\documentclass{aa}
\usepackage{natbib}
\usepackage{epsfig}
\usepackage{graphicx}
\usepackage{xspace}
\newcommand{\MURaM}{\texttt{MURaM}\xspace}
\newcommand{\COBOLD}{\texttt{CO$^5$BOLD}\xspace}
\newcommand{\STAGGER}{\texttt{Stagger}\xspace}
\begin{document}
\title{Simulations of the solar near-surface layers with the
       CO5BOLD, MURaM, and Stagger codes}
\author{B. Beeck\inst{1}, R. Collet\inst{2}, M. Steffen\inst{3},
        M. Asplund\inst{2,4}, R.H. Cameron\inst{1}, B. Freytag\inst{5}, 
        W. Hayek\inst{6},
        H.-G. Ludwig\inst{7}, \and M. Sch\"ussler\inst{1} 
        }
\institute{Max-Planck-Institut f\"ur Sonnensystemforschung,
           37191 Katlenburg-Lindau, Germany
	   \and
	   Max-Planck-Institut f\"ur Astrophysik, 
           Karl-Schwarzschild-Str. 1, 85741 Garching, Germany,
	   \and
           Leibniz-Institut f\"ur Astrophysik Potsdam (AIP), 
	   An der Sternwarte 16, 14482 Potsdam, Germany
	   \and
	   Research School of Astronomy and Astrophysics,
	   Australian National University, 
	   Cotter Rd, Weston Creek, ACT 2611, Australia
	   \and
	   Centre de Recherche Astrophysique de Lyon,
            UMR 5574: CNRS, Universit\'e de Lyon,
            \'Ecole Normale Sup\'erieure de Lyon,
            46 All\'ee d'Italie, 69364 Lyon Cedex 07, France
	   \and
           School of Physics, University of Exeter, Stocker Road,
           Exeter EX4 4QL, UK
           \and
	   ZAH, Landessternwarte, K{\"o}nigstuhl 12, 
           69117 Heidelberg, Germany}
\date{\today}
\abstract
  {Radiative hydrodynamic simulations of solar and stellar surface
  convection have become an important tool for exploring the structure
  and gas dynamics in the envelopes and atmospheres of late-type stars
  and for improving our understanding of the formation of
  stellar spectra. }
  {We quantitatively compare results from three-dimensional, radiative
  hydrodynamic simulations of convection near the solar surface
  generated with three numerical codes (\COBOLD, \MURaM, and \STAGGER)
  and different simulation setups in order to investigate the level of
  similarity and to cross-validate the simulations.}
  {For all three simulations, we considered the average stratifications
   of various quantities (temperature, pressure, flow velocity, etc.)
   on surfaces of constant geometrical or optical depth, as well as
   their temporal and spatial fluctuations.  We also compared
   observables, such as the spatially resolved patterns of the emerging
   intensity and of the vertical velocity at the solar optical surface
   as well as the center-to-limb variation of the continuum intensity at
   various wavelengths.}
  {The depth profiles of the thermodynamical quantities and of the
  convective velocities as well as their spatial fluctuations agree
  quite well. Slight deviations can be understood in terms of
  differences in box size, spatial resolution and in the treatment of
  non-gray radiative transfer between the simulations.}
  {The results give confidence in the reliability of the results from
    comprehensive radiative hydrodynamic simulations.}
    \keywords{Methods: numerical -- Sun: photosphere -- convection --
    hydrodynamics -- radiative transfer }
\authorrunning{Beeck et al.}
\titlerunning{Simulations of the solar near-surface layers}
\maketitle
\section{Introduction}
Comprehensive (magneto)hydrodynamic simulations have become an essential
tool for studying near-surface convection in the Sun and other cool
stars, together with the structure and gas dynamics in their atmospheres
\citep[e.g.,][]{Nordlund:etal:2009}.  These simulations attempt to
include all relevant physics, such as three-dimensional (3D),
time-dependent, compressible hydrodynamics, partial ionization and
molecule formation as well as non-gray and non-local radiative transfer,
in order to provide a `realistic' representation of the physical
stratification and macroscopic gas flows in the external stellar layers.
For a direct comparison with observations, spectral line profiles,
continuum intensity and polarization maps are calculated on the basis of
the simulation results.  This comparison serves as a means of validation
of the simulations and also as a tool for interpreting the observational
results in terms of basic physical quantities
\citep[cf.][]{Uitenbroek:Criscuoli:2011}.

\begin{table*}[ht!]
\caption{Numerical methods$^\mathrm{a}$ used in the codes.}
\label{tab:methods}
\begin{tabular}{c c c c c c}
\hline
\noalign{\vspace{2mm}}
Code & spatial scheme& temporal scheme & RT scheme &\# rays & \# bins\\[0.5ex]
\hline
\rule[3mm]{0mm}{2mm}
\COBOLD & \multicolumn{2}{c}{Roe-type Riemann} &
          long characteristics & 17 & 12\\ [0.5ex]
\rule[3mm]{0mm}{2mm}
\MURaM  & $4^{\rm th}$-order FD & $4^{\rm th}$-order RK &
          short characteristics & 12 & 4\\ [0.5ex]
\rule[3mm]{0mm}{2mm}
\STAGGER& $6^{\rm th}$-order FD & $3^{\rm th}$-order RK &
          long characteristics & 9 & 12\\ [0.5ex]
\hline
\end{tabular}
\begin{list}{}{}
\item[$^\mathrm{a}$]  FD: finite differences,
 RK: Runge-Kutta, RT: radiative transfer.
\end{list}
\end{table*}

Although various codes are now being used to perform comprehensive
simulations of solar and stellar (magneto)convection and an extensive
body of simulation results has already been published, so far no
systematic attempt has been made to cross-validate codes by
quantitatively comparing numerical results.  In this paper, we attempt
to fill this gap, at least partially, and compare the solar models
computed with \COBOLD, \MURaM, and \STAGGER, three independent and
widely used 3D, radiative (magneto)hydrodynamic simulation codes.  Apart
from these codes, a number of other codes for 3D simulations of solar
and stellar surface convection including (full or simplified) radiative
transfer have been developed and utilized by various groups
\citep[e.g.,][]{Stein:Nordlund:1998, Robinson:etal:2003,
Heinemann:etal:2006, Abbett:2007, Ustyugov:2009, Muthsam:etal:2010,
Gudiksen:etal:2011}.

The purpose of our study is not a comparison of the numerical approaches
of \COBOLD, \MURaM, and \STAGGER per se, which would require using an
identical setup in terms of box size, spatial resolution, and input
material quantities such as opacities and equation of state. We rather
wish to investigate how far simulations made for different applications
are consistent in the basic properties of the simulated stellar
atmosphere and uppermost convection zone. Examples of such properties
are the average profiles of various quantities as a function of
geometrical or optical depth. According to this rationale, we chose for
comparison `standard' simulations of the near-surface layers of the Sun
that were carried out by the participating groups for different
purposes. The \COBOLD and \STAGGER simulations provide a solar reference
atmosphere (as part of a large grid of stellar models) for
spectrum-synthesis calculations and abundance studies; they thus focus
upon a good representation of the energy exchange by non-gray radiative
transfer. The \MURaM simulation, on the other hand, represents a
non-magnetic comparison model for a set of magnetoconvection simulations
to study fine-scale magnetic phenomena, for which high spatial
resolution is crucial. The question we address here is: how much do the
basic properties of the `solar models' resulting from simulations with
different codes, different input quantities, different setup in terms of
box size and spatial resolution, and even different top boundary
conditions deviate from each other? If the differences turn out to be
marginal, this result could then be taken as a kind of
`cross-validation' of the codes and as a basis for confidence in the
reliability of the simulations.

\section{Codes}

All three codes considered here treat the coupled time-dependent
equations of compressible radiative (magneto)hydro\-dynamics and
radiative transfer in a three-dimensional geometry and for a stratified,
partially ionized medium.  The energy exchange between radiation and
matter is accounted for through solving the equation of radiative
transfer under the assumption of local thermodynamic equilibrium (LTE)
with a Planckian source function.  To reduce the computational load, the
wavelength-dependence of the radiative transfer is treated with the
method of opacity binning \citep{Nordlund:1982, Ludwig:1992,
Skartlien:2000, Voegler:etal:2004}.  A brief overview of the numerical
methods used is given in Table~\ref{tab:methods}. In all cases, a local
`box-in-the-star' setup is employed: the computational domain is a
rectangular 3D box straddling in height the photosphere and the
uppermost few Mm of the convection zone.%
  \footnote{However, by number of scale heights, the simulations cover
  about a third of the total pressure range of the convection zone.}
The simulation boxes are sufficiently extended in the horizontal
directions (6--9~Mm) to contain about 15--40 convection cells (granules)
at any given time, thus providing a statistically useful sample of the
near-surface layers of the Sun. Periodic boundary conditions are assumed
in the horizontal directions (side boundaries) while the bottom boundary
is open and allows free in- and outflow of fluid. A fixed entropy
density is prescribed for the inflowing fluid at the lower boundary. It
can be interpreted as the entropy of the deep, almost adiabatically
stratified convective envelope and controls the effective temperature of
the simulated atmosphere. In addition, the gas pressure is kept constant
across the bottom boundary.  The three codes differ somewhat in their
treatment of the upper boundary conditions, as outlined more
specifically below.

Results from all three codes considered here already passed various
`reality checks' by comparison with observational data
\citep[e.g.,][]{Danilovic:etal:2008, Pereira:etal:2009b,
Pereira:etal:2009a, Wedemeyer:etal:2009, Hirzberger:etal:2010}.

\subsection{CO$^5$BOLD}

\begin{table}[h!] 
\caption{Optical depth ranges and wavelength
ranges$^\mathrm{a}$ of the opacity bins.}
\begin{tabular}{r r r r r}\hline
\noalign{\vspace{1mm}} 
\multicolumn{5}{c}{\COBOLD}\\ \hline
\noalign{\vspace{1mm}} 
Bin & $\log\tau_1$ & $\log\tau_2$ & $\lambda_1$ [nm] & $\lambda_2$ [nm] 
     \\[0.5ex]\hline
1 &   0.15  &  99.00 &    0.  &     550.\\ [0.0ex] 
2 &   0.15  &  99.00 &  550.  &  100000.\\ [0.0ex] 
3 &   0.00  &   0.15 &    0.  &     600.\\ [0.0ex] 
4 &   0.00  &   0.15 &  600.  &  100000.\\ [0.0ex] 
5 &  $-$0.75  &   0.00 &    0.  &     650.\\ [0.0ex] 
6 &  $-$0.75  &   0.00 &  650.  &  100000.\\ [0.0ex] 
7 &  $-$1.50  &  $-$0.75 &    0.  &  100000.\\ [0.0ex] 
8 &  $-$2.25  &  $-$1.50 &    0.  &  100000.\\ [0.0ex] 
9 &  $-$3.00  &  $-$2.25 &    0.  &  100000.\\ [0.0ex] 
10 &  $-$3.75 &   $-$3.00 &   0.  &  100000.\\ [0.0ex] 
11 &  $-$4.50 &   $-$3.75 &   0.  &  100000.\\ [0.0ex] 
12 & $-$99.00 &   $-$4.50 &   0.  &  100000.\\ [0.0ex] 
\hline
\noalign{\vspace{1mm}} 
\multicolumn{5}{c}{\MURaM}\\ [0.0ex] \hline
\noalign{\vspace{1mm}} 
1 &   0.00  &  99.   &   &    \\ [0.0ex] 
2 &  $-$2.00  &  0.00  &   &    \\ [0.0ex] 
3 &  $-$4.00  & $-$2.00  &   &    \\ [0.0ex] 
4 &  $-$99.   & $-$4.00  &   &    \\ [0.0ex] 
\hline
\noalign{\vspace{1mm}} 
\multicolumn{5}{c}{\STAGGER}\\ [0.0ex] \hline
\noalign{\vspace{1mm}} 
1 & $-$1.46  &   9.00  &     0.  &   380.9\\ [0.0ex] 
2 & $-$3.81  &  $-$1.46  &     0.  &   380.9\\ [0.0ex] 
3 & $-$15.00 &  $-$3.81  &     0.  &   380.9\\ [0.0ex] 
4 & $-$0.62  &   9.00  &  380.9  &   562.4\\ [0.0ex] 
5 & $-$0.62  &   9.00  &  562.4  &  2161.2\\ [0.0ex] 
6 & $-$1.50  &  $-$0.62  &  380.9  &   642.6\\ [0.0ex] 
7 & $-$2.28  &  $-$1.50  &  380.9  &   710.9\\ [0.0ex] 
8 & $-$10.00 &  $-$2.28  &  380.9  &  1646.5\\ [0.0ex] 
9 & $-$0.62  &   9.00  &  2161.2 &  100000.\\ [0.0ex] 
10& $-$1.50  &  $-$0.62  &  642.6  &  100000.\\ [0.0ex] 
11& $-$2.28  &  $-$1.50  &  710.9  &  100000.\\ [0.0ex] 
12& $-$10.00  & $-$2.28  & 1646.5  &  100000.\\ [0.0ex] 
\hline
\end{tabular}
\begin{list}{}{}
\item[$^\mathrm{a}$]  The bins in the \MURaM simulation are
wavelength-independent.
\end{list}
\label{tab:binning}
\end{table}

The \COBOLD code uses a numerical scheme based on a finite-volume
approach on a fixed Cartesian grid. Operator splitting separates the
various (usually explicit) operators: the (magneto)hydrodynamics, the
tensor viscosity, the radiation transport, and optional source
steps. Directional splitting reduces the multi-dimensional hydrodynamics
problem to a sequence of 1D steps. The advection step is performed by an
approximate Riemann solver of Roe type, modified to account for a
realistic equation of state, a non-equidistant grid, and the presence of
source terms due to an external gravity field.  Optionally, a 3D tensor
viscosity can be activated for improved stability in extreme situations.
Parallelization of \COBOLD is achieved with OpenMP.

The top boundary condition provides transmission of waves of arbitrary
amplitude, including shocks: typically, two layers of ghost cells are
introduced, where the velocity components and the internal energy are
kept constant and the density decreases exponentially with a scale
height set to a controllable fraction of the local hydrostatic pressure
scale height. This gives the possibility to minimize the mean mass flux
through the open top boundary.

The radiative transfer in the \COBOLD simulation considered here was
computed using a long-characteristics scheme for rays with four
inclination angles and four azimuthal angles plus the vertical, i.e. 17
rays in total. The values of the inclinations ($\mu=\cos\theta$=$1.000$,
$0.920$, $0.739$, $0.478$, $0.165$) correspond to the positive nodes of
the $10^\mathrm{th}$-order Lobatto quadrature formula, the four
azimuthal angles coincide with the grid directions ($\phi$=$0$, $\pi/2$,
$\pi$, and $3\pi/2$).  For the opacity binning, tables were constructed
from a data set of MARCS raw opacities provided by B. Plez
\citep[priv. comm.; see also][]{Gustafsson:etal:2008}, comprising
continuous and sampled atomic and molecular line opacities as functions
of temperature and gas pressure at more than $10^5$~wavelength
points. The adopted chemical composition comes from
\citet{Asplund:etal:2005}. Each wavelength of the original opacity
sampling data was sorted into one of twelve representative bins,
according to wavelength and Rosseland optical depth where the
monochromatic optical depth unity is reached in a 1D standard solar
atmosphere. The thresholds for the opacity bins used for the present
\COBOLD solar simulation are given in Table~\ref{tab:binning}.  For each
opacity bin, the tabulated opacity is a hybrid of Rosseland and Planck
means over all frequencies of the bin, such that it approaches the
Rosseland mean at high values of the optical depth, and the Planck mean
at low values, with a smooth transition centered at Rosseland
optical depth $0.35$. Assuming LTE and pure absorption, the source
function in each bin is computed as the Planck function integrated over
the frequencies associated with the respective bin.

The equation of state (EOS) used in \COBOLD follows \citet{Wolf:1983}.
It accounts for the partial ionization of hydrogen and helium, as well
as for H$_2$ molecule formation.  In contrast to Wolf's approach, all
pressure-temperature regions are treated homogeneously since performance
optimization was not necessary because a tabulated EOS is used.

The \COBOLD simulations considered in this paper was used by
\citet{Caffau:etal:2008} for the determination of the solar thorium and
hafnium abundances, and for subsequent studies of CNO and other
elements. More details about the \COBOLD code can be found in
\citet{Freytag:etal:2002}, \citet{Wedemeyer:etal:2004}, and
\citet{Freytag:etal:2008,Freytag:etal:2011}.

\subsection{MURaM}

The \MURaM code \citep{Voegler:2003, Voegler:etal:2005} uses a
$4^\mathrm{th}$-order central difference scheme in space and a
$4^\mathrm{th}$-order Runge-Kutta scheme for time-stepping. Artificial
diffusivities are treated with the scheme described in
\citet{Rempel:etal:2009}.  An open-top boundary condition is also
implemented in the \MURaM code, but for the simulation considered here a
stress-free, closed top (zero vertical velocity) was chosen in
order to study how far this affects the mean stratification.  \MURaM
uses the Message Passing Interface (MPI) framework for parallelization.

Radiative transfer in the \MURaM code is calculated with the
short-characteristics method \citep{Kunasz:Auer:1988} with bilinear
interpolation. The angular integration is carried out according to the
A4 scheme of \citet{Carlson:1963} along three directions per octant, which
corresponds to 12 complete rays in total \citep[cf.][]{Bruls:etal:1999}. The
opacity binning for the non-gray radiative transfer is based on the
opacity distribution functions from the ATLAS9 package
\citep{Kurucz:1993} and uses 4 bins. The thresholds in optical depth
(see Table~\ref{tab:binning}) for the binning procedure were chosen in
terms of $\log\overline{\tau}$, which is a hybrid of the Rosseland mean
in the deeper layers and the Planck mean in the upper layers, with a
smooth transition centered at $\tau=0.35$ \citep{Ludwig:1992,
Voegler:etal:2004}. The EOS tables used for the simulation considered
here are based on tables from the OPAL project \citep{Rogers:etal:1996}
for a solar gas mixture with abundances from
\citet{Anders:Grevesse:1989}.

\subsection{Stagger}

The \STAGGER code, \citep[originally developed
by][]{Galsgaard:Nordlund:1996}%
\footnote{see also
  \texttt{http://www.astro.ku.dk/$\sim$kg/Papers/MHD\_code.ps.gz}}%
, uses a $6^\mathrm{th}$-order finite
difference scheme in space with $5^\mathrm{th}$-order interpolations.
Scalar variables (density, internal energy, and temperature) are
volume-centered, while momenta are face-centered. The hydrodynamic
variables are advanced forward in time using a $3^\mathrm{rd}$-order
Runge-Kutta scheme.  Boundaries are periodic horizontally and open
vertically, both at the top and at the bottom.  The EOS is taken from
\citet{Mihalas:etal:1988} and accounts for the effects of excitation,
ionization, and dissociation of the 15 most abundant elements and of the
$H_2$ and $H_2^+$ molecules. Parallelization of the \STAGGER code is
carried out via MPI.

The radiative-transfer equation is solved with a Feautrier-like
\citep[][]{Feautrier:1964} scheme along eight inclined rays (two
inclination angles, four azimuth angles) plus the vertical, and using an
opacity-binning scheme with twelve bins for the frequency dependence.
The total radiative heating rate at the center of each grid cell is
computed by adding the partial contributions from each direction and
opacity bin with the appropriate weight.  The values of the inclination
angles ($\mu$=$\cos\theta$=$0.155$, $0.645$, $1.000$) and their
associated weights ($w_\mu$=$0.376$, $0.512$, $0.111$) correspond to the
nodes and weights of the $3^\mathrm{rd}$-order Radau quadrature formula;
the four azimuthal angles are equidistant ($\phi$=$0$, $\pi/2$, $\pi$,
and $3\pi/2$) and have equal weights.

\begin{table}[h!] 
\caption{Parameters of the simulation runs.}
\begin{tabular}{c c c c }
\hline
\noalign{\vspace{1mm}} 
Code & Box [Mm$^3$] & $h$ [Mm]$^\mathrm{b}$ & grid resolution [km]
     \\
     & $(x,y,z)^\mathrm{a}$     &  & $(x,y,z)$  
     \\[0.5ex]
\hline
\rule[3mm]{0mm}{2mm} 
\COBOLD & $5.6 \times 5.6  \times 2.3$ & 0.88  & $40\times 40\times 15.1$ 
       \\ [0.5ex] 
\rule[3mm]{0mm}{2mm} 
\MURaM  & $9   \times 9    \times 3 $ & 1 & $17.6\times 17.6\times 10$ 
       \\ [0.5ex]      
\rule[3mm]{0mm}{2mm} 
\STAGGER& $6\times 6  \times 3.6$ & 0.88 & $25.1\times 25.1\times 7...32$
       \\ [0.5ex]    
\hline
\end{tabular}
\begin{list}{}{}
\item[$^\mathrm{a}$] $x,y$: coordinates in the horizontal directions;\\ 
                     $z$:  depth coordinate.
\item[$^\mathrm{b}$] height of top boundary above $\tau=1$   
\end{list}
\label{tab:parameters}
\end{table}
\begin{table}[ht!] 
\caption{Global properties of the simulated solar models.}
\begin{tabular}{c c c c}
\hline
\noalign{\vspace{0.5mm}} 
Code & $T_{\rm eff} [K] $ 
     & $(\delta I/I)_{\mathrm bol}[\%]$ 
     & $(\delta I/I)_{\mathrm 500}[\%]$ 
     \\[0.5ex]
\hline
\rule[3mm]{0mm}{2mm} 
\COBOLD & $5782.1\pm12.6$ & $14.4\pm0.6$  & $21.8\pm0.8$ \\ [0.5ex] 
\rule[3mm]{0mm}{2mm} 
\MURaM  & $5768.4\pm9.9$\ \ \ & $15.4\pm0.3$  & $21.8\pm0.3$ \\ [0.5ex]      
\rule[3mm]{0mm}{2mm} 
\STAGGER & $5778.4\pm15.8$ & $15.1\pm0.5$ & $22.1\pm0.8$ \\ [0.5ex]    
\hline
\end{tabular}
\label{tab:properties}
\end{table}

For the \STAGGER simulation considered here, the continuous opacity data
came from \citet{Gustafsson:etal:1975} and Trampedach (private
communication), while sampled line opacities were taken from the {\sc
MARCS} package \citep[B. Plez, private communication; see
also][]{Gustafsson:etal:2008}. The adopted chemical composition for the
simulation considered here was taken from \citet{Asplund:etal:2005}.

The opacity binning procedure implemented in the \STAGGER code is
essentially based on the formulation by
\citet{Skartlien:2000}. Opacities are sorted into bins according to
their wavelength and strength.  As a measure of the opacity strength at
a given wavelength, the Rosseland optical depth of formation of that
particular wavelength is used.  More precisely, the formation depth is
defined as the point where the monochromatic optical depth in the
vertical direction equals unity in a one-dimensional model constructed
by taking the mean temperature-density stratification from a solar
simulation.  The thresholds in Rosseland optical depth and wavelength
for the determination of bin membership are given in
Table~\ref{tab:binning}.  Within each opacity bin, opacities are
averaged and the source function contributions at the various
wavelengths belonging to the bin are integrated. Mean-intensity-weighted
average opacities and Rosseland-like mean opacities are adopted in the
optically thin and optically thick layers, respectively. A bridging
function is used for a smooth transition between the two averages near
the optical surface.  In the \STAGGER simulation considered here, the
Planck function at the local temperature was chosen as the source function
and the contribution of scattering to the total opacity in the optically
thin layers was neglected. The simulation belongs to the series that was
used in the recent analysis by \citet{Asplund:etal:2009} for the
spectroscopic determination of solar abundances.  A more comprehensive
description of the opacity binning implementation and of the
approximations involved in current \STAGGER code simulations are given
in \citet{Collet:etal:2011}.

\section{Simulation runs and quantities for comparison}
The three codes were used to carry out simulations of near-surface solar
convection without magnetic field. As explained in the introduction, the
simulation setups were different, corresponding to the different research
topics that the participating groups focus upon. Therefore, the
numerical setups differ in terms of (horizontal and vertical) box size,
grid resolution, number of opacity bins, and other features such as the
top boundary condition.  In all cases, the simulation boxes include the
photosphere (up to about $1$~Mm above the optical surface) and the
uppermost layers of the convection zone (between $1.4$~Mm and $3$~Mm
below the optical surface, depending on the
simulation). Table~\ref{tab:parameters} gives various parameters of the
simulation runs. Note that the \STAGGER code uses a non-equidistant grid
of 230 cells in the vertical direction, with spacings ranging from 7~km
around the optical surface and 32~km in the deepest parts of the
simulation box.

The simulations were run for several hours of solar time to reach a
statistically stationary, thermally relaxed state. Nineteen snapshots
taken at regular intervals and spanning in total a period of about two
hours were considered for the analysis of each simulation. This choice
was made to ensure that the effects of the 5-minute $p$-mode
oscillations in the simulations are averaged out in the temporal
means.  

\begin{figure*}[ht!]
\vspace*{0mm}
\centering
  \resizebox{1.0\hsize}{!}{\includegraphics{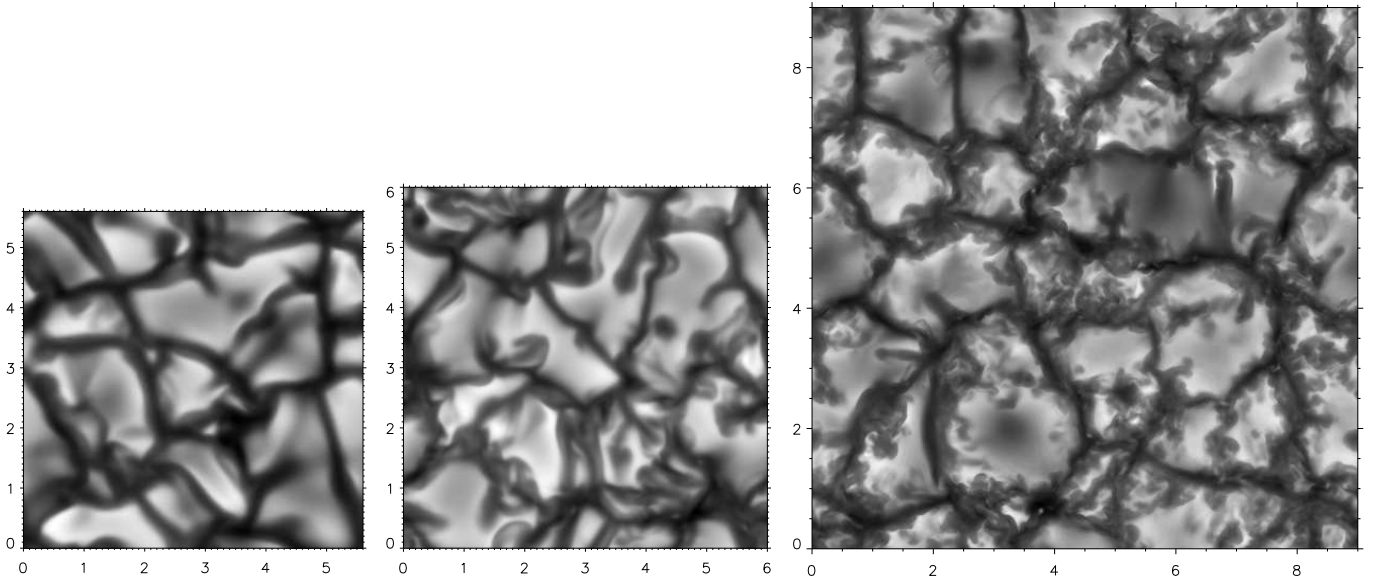}}
\caption{Vertically emerging continuum intensity at 500~nm for single
  snapshots from the \COBOLD (left), \STAGGER (middle), and \MURaM
  (right) runs, drawn to scale. The gray scales cover, from black to
  white, the ranges 0.59--1.53 (\COBOLD), 0.50--1.53 (\STAGGER), and
  0.49--1.66 (\MURaM) of the intensity normalized to the respective
  horizontal average. Axis units are Mm.}
\label{fig:i_500}
\end{figure*}
\begin{figure*}[ht!]
\vspace*{0mm}
\centering
  \resizebox{1.0\hsize}{!}{\includegraphics{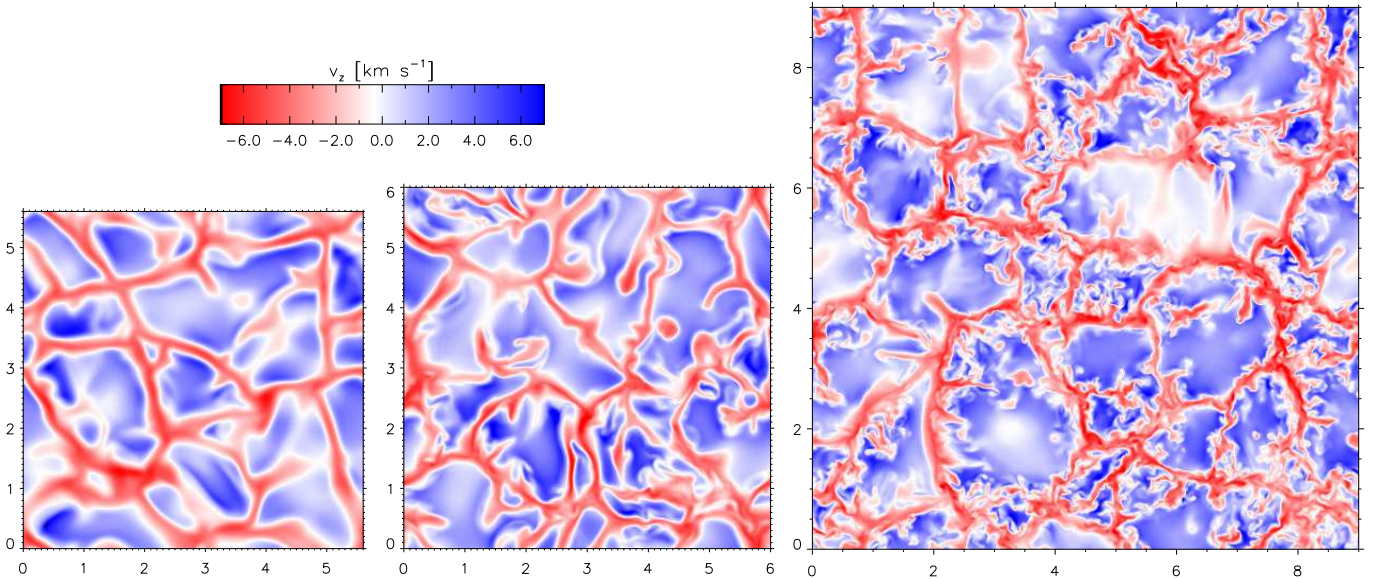}}
\caption{Vertical velocity at the average geometrical depth level
of the surface $\tau_{500}=1$ for the same snapshots as in
Fig.~\ref{fig:i_500}. Downflows are shown in red, upflows in blue. The
color table covers the range $\pm 7\,$km$\,\,$s$^{-1}$ in all cases;
speeds outside this range are saturated. Axis units are Mm.}
\label{fig:vz_1}
\end{figure*}

The physical quantities considered for the comparison are temperature,
gas pressure, and turbulent pressure ($\rho v_z^2$), as well as the
vertical and horizontal velocity components. To obtain mean 
profiles as functions of depth, $z$, for each of these quantities,
$q(x_i,y_j,z_k,t_l)\equiv q_{ijk,l}$, and their squares at the grid
cells $(x_i,y_j,z_k)$ and at time $t=t_l$, the averages over horizontal
planes ($z_k={\rm const.}$) were determined, viz.
\begin{eqnarray}
  \overline{q}_{k,l} 
   &=& \frac{1}{n_x n_y}\sum_{i=1}^{n_x}\sum_{j=1}^{n_y}q_{ijk,l}\\
  \overline{q^2}_{k,l} 
   &=& \frac{1}{n_x n_y}\sum_{i=1}^{n_x}\sum_{j=1}^{n_y}q^2_{ijk,l}\;,
\end{eqnarray}
where $n_x$ and $n_y$ are the number of grid cells in the horizontal
directions. Similarly, averages over surfaces of constant optical depth
were determined by first calculating the optical depth along vertical
lines of sight, $\tau_{500}$, for the continuum opacity at 500~nm
wavelength and then considering the quantities at fixed levels in the
range $-4\leq\log\tau_{500}\leq 4$. We also considered averages based on
optical depth corresponding to the Rosseland mean opacity and found the
results to be not significantly different from those with 500~nm
continuum opacity, so that we restrict ourselves to the latter case.

From the vertical profiles $\overline{q}_{k,l}$ and $\overline{q^2}_{k,l}$
we determine temporal averages over the $N=19$ snapshots considered for
each simulation,
\begin{equation}
  \langle \overline{q} \rangle_k = 
        \frac{1}{N}\sum_{l=1}^{N}\overline{q}_{k,l}\;,
\label{eq_mean}
\end{equation}
the standard deviation among the snapshots,
\begin{equation}
  \sigma(\overline{q})_k = \left[ \frac{1}{N}\sum_{l=1}^{N}
                \left( \overline{q}_{k,l}-\langle \overline{q} \rangle_k 
                \right)^2 \right]^{1/2} ,
\label{eq_stddev}
\end{equation}
and the temporal average of the spatial root-mean-square (RMS)
fluctuation,
\begin{equation}
  \langle q_{\rm RMS}\rangle_k = \frac{1}{N}\sum_{l=1}^{N}
                \left[ \overline{q^2}_{k,l}
                -\left(\overline{q}_{k,l}\right)^2
                \right]^{1/2} .
\label{eq_rms}
\end{equation}
The index $k$ in Eqs.~(\ref{eq_mean}--\ref{eq_rms}) refers either to 
the geometrical depth level, $z_k$, for averages over planes of constant
geometrical depth or to the optical depth level, $\tau_{500,k}$, for
averages over surfaces of constant optical depth. Strictly speaking,
the latter averages are taken over the projections of the
corrugated surfaces of constant optical depth on a horizontal plane.

\section{Results}

Table~\ref{tab:properties} shows the effective temperatures of the
various simulation models together with the disk-center bolometric and
monochromatic continuum intensity contrasts at 500~nm. The standard
deviations indicate the variability among the 19 snapshots from each
simulation run used in the analysis. The effective temperatures differ
by about 14~K at most and the intensity contrasts agree fairly well with
each other.  In comparison to the other simulations, the variations from
snapshot to snapshot are smaller in the \MURaM case. This is probably
because of the larger horizontal extension of the computational box.

Unless stated otherwise, all quantities discussed in the following
subsections refer to averages over the 19 snapshots from each of the
simulation runs.

\begin{figure*}[ht!]
\vspace*{0mm}
\centering
  \resizebox{0.95\hsize}{!}{\includegraphics{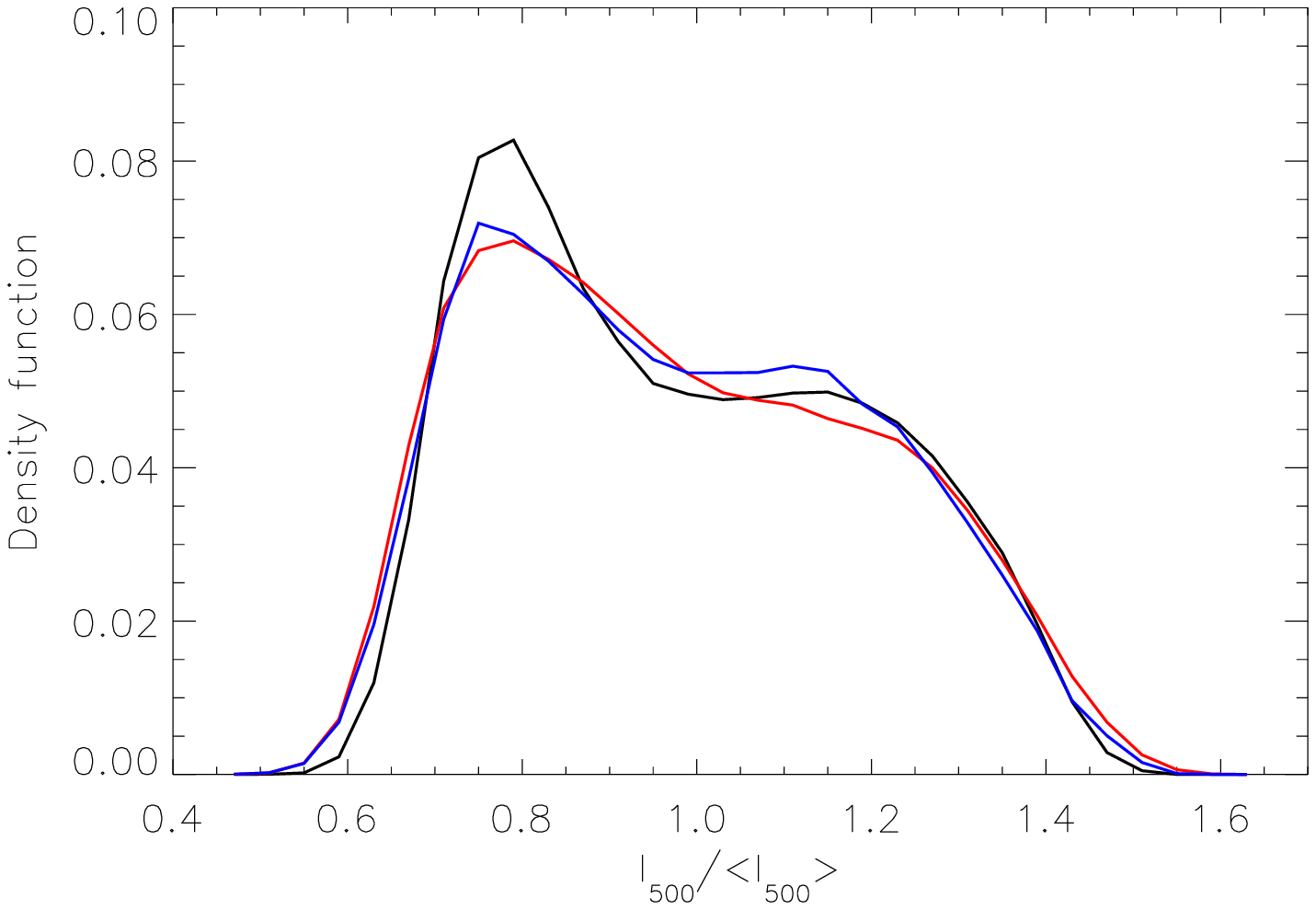}
                          \includegraphics{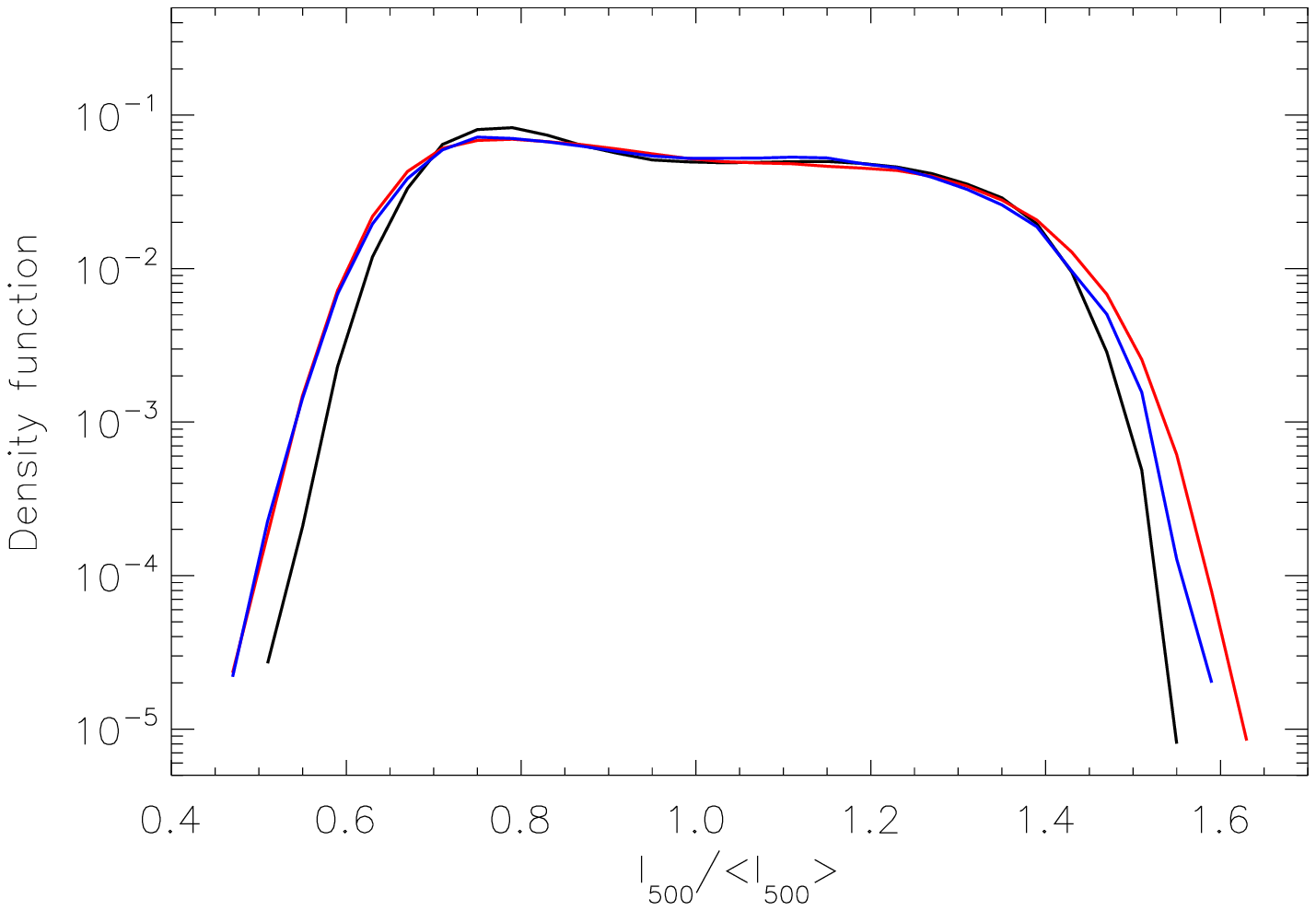}}
  \resizebox{0.95\hsize}{!}{\includegraphics{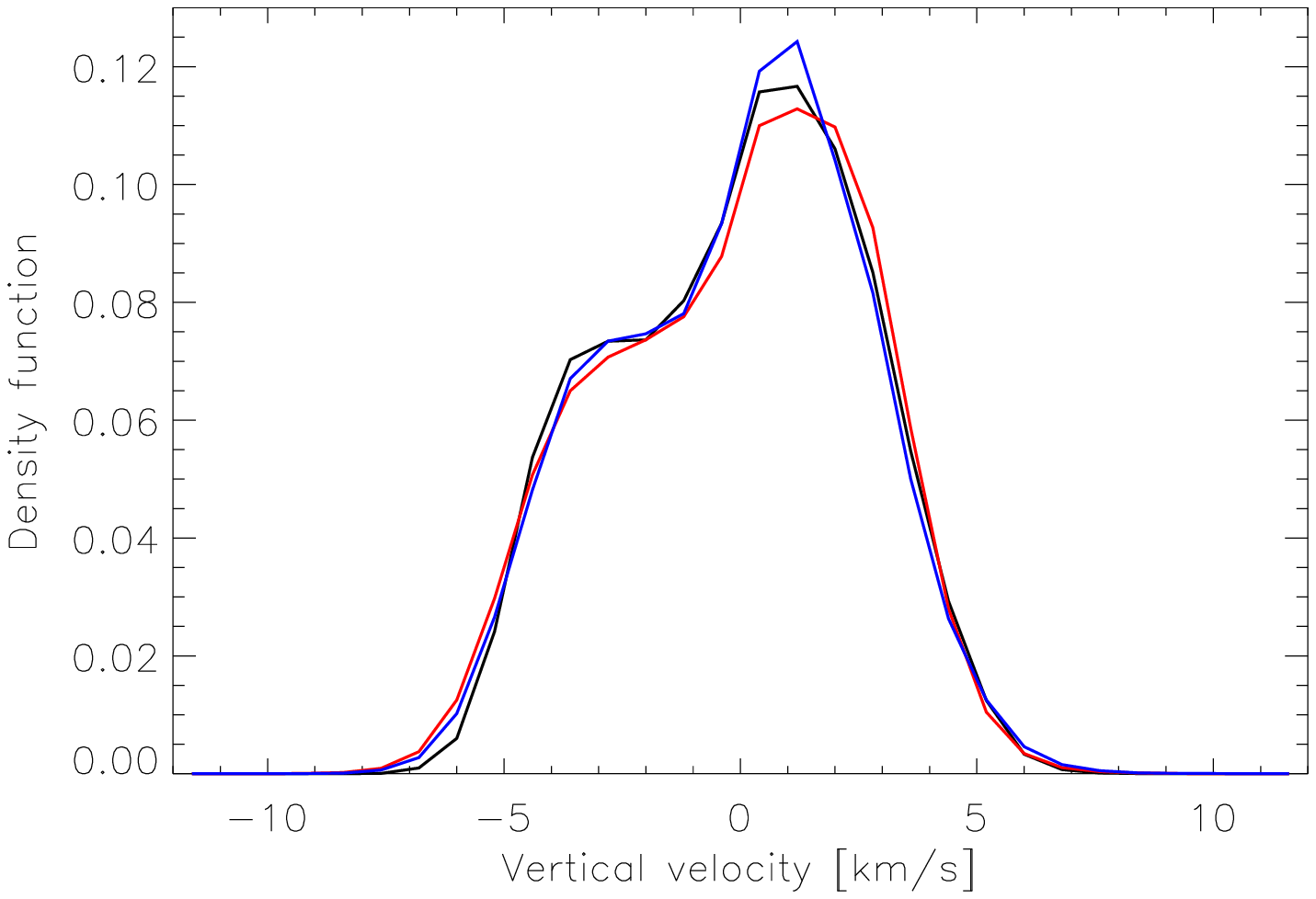}
                          \includegraphics{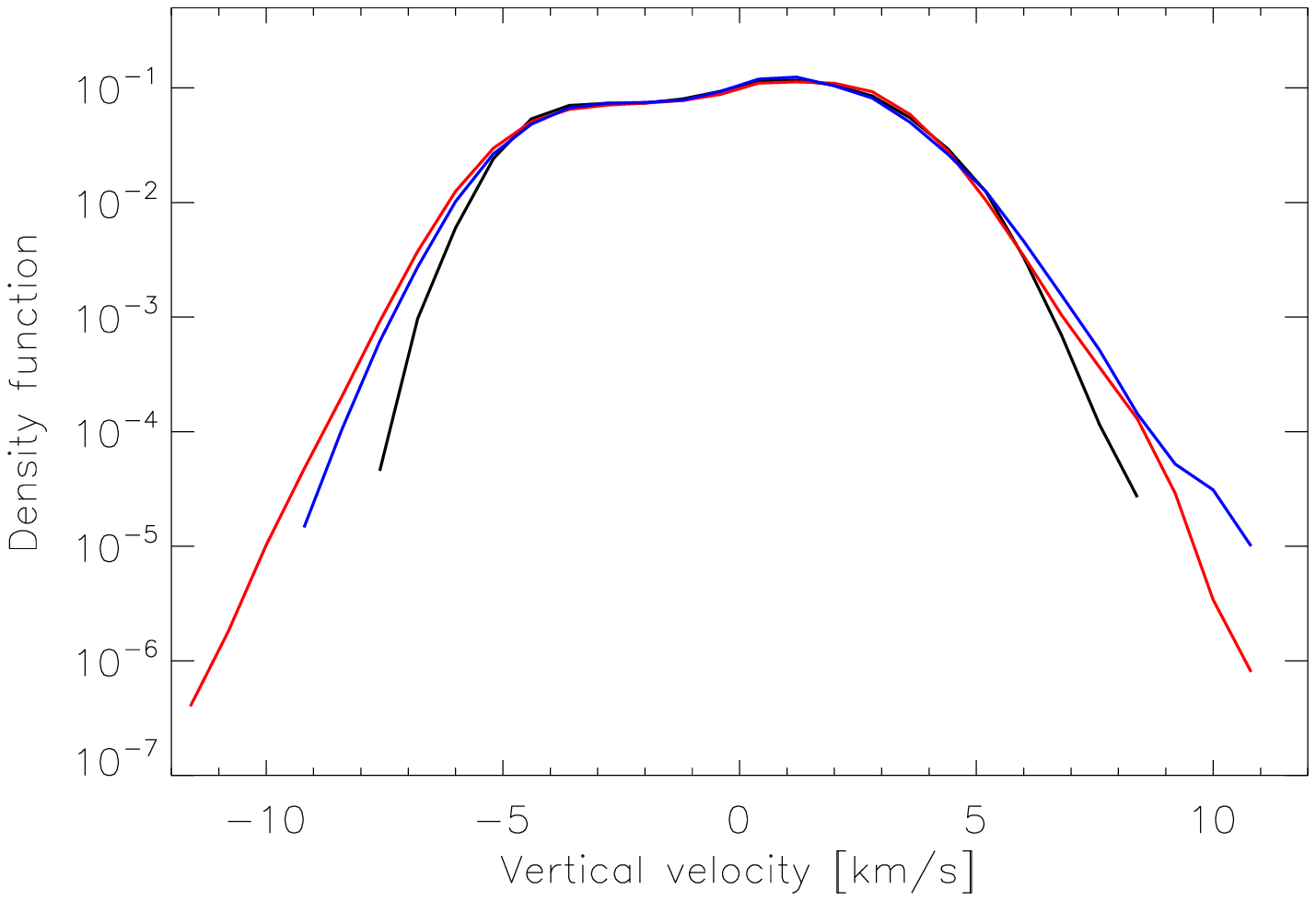}}
\caption{Linear (left) and logarithmic (right) histograms of the
  vertically emerging continuum intensity at 500~nm (upper panels) and
  the vertical velocity on the average height level of the surface
  $\tau_{500}=1$ (lower panels) averaged over all 19 snapshots from each
  simulation (black: \COBOLD, red: \MURaM, blue: \STAGGER). Positive
  velocities correspond to upflows. Thirty bins were used in all
  cases. Each histogram was normalized such that the sum of the density
  function over the bins becomes unity.}
\label{fig:histograms}
\end{figure*}

\subsection{Surface maps and histograms}

Figure~\ref{fig:i_500} shows maps of the vertically emerging
(disk-center) continuum intensity at 500~nm for snapshots from the three
simulation runs. Fig.~\ref{fig:vz_1} gives the corresponding maps of the
vertical velocity at the average geometrical depth level of the surface
$\tau_{500}=1$, where $\tau_{500}$ is the continuum optical depth at
500~nm wavelength. The runs with higher spatial resolution naturally
show more small-scale details, but the basic structure and average size
of the granules and the correlation between the brightness and velocity
are very similar in all simulations. The visual impression is confirmed
by the similarity of the histograms of intensity and vertical velocity
given in Fig.~\ref{fig:histograms}. On a logarithmic scale (right
panels), the difference in spatial resolution becomes apparent at the
extreme values, but otherwise there are no significant differences
between the distributions.

\begin{figure*}[ht!]
\vspace*{0mm}
\centering
  \resizebox{0.8\hsize}{!}{\includegraphics{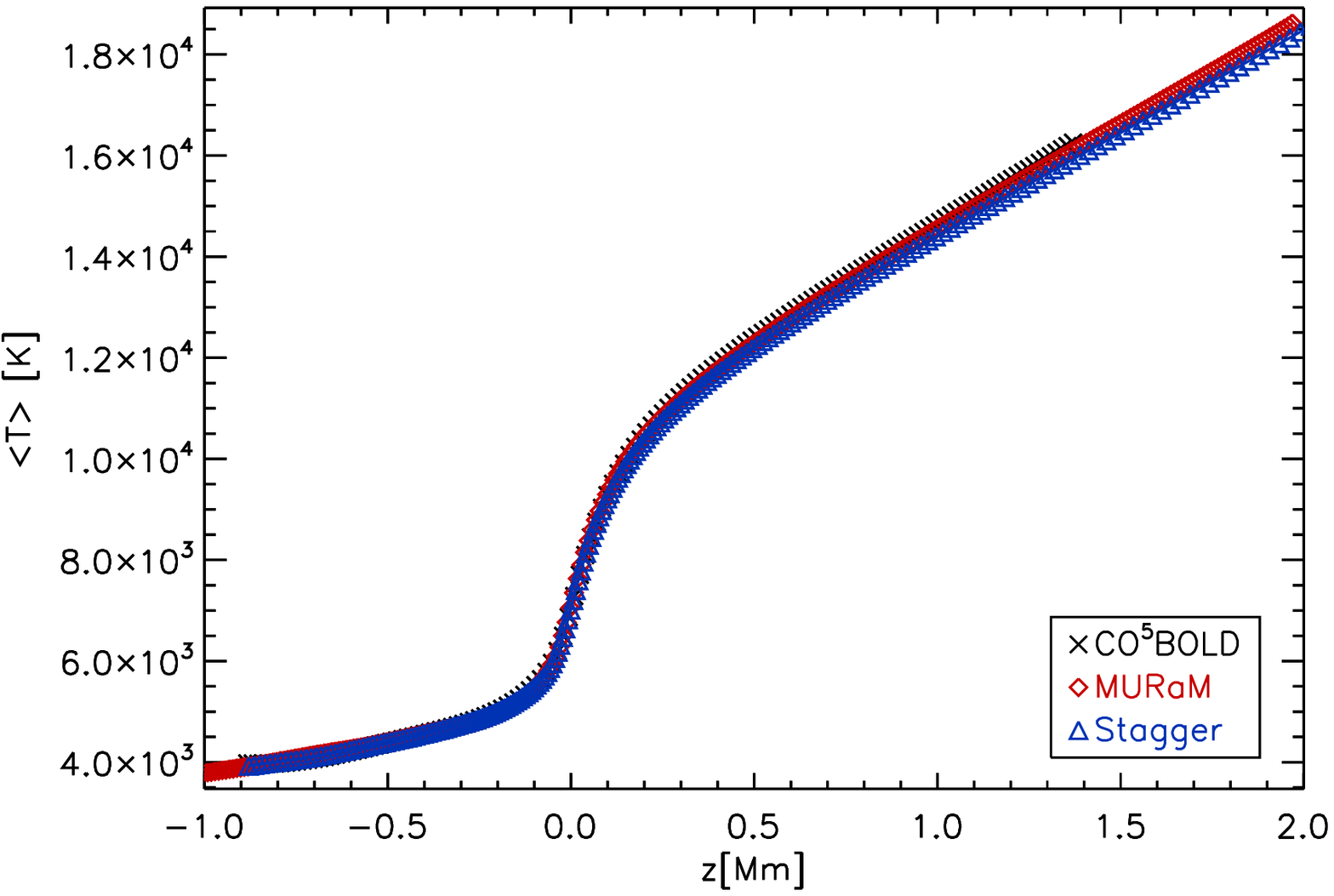}
                          \includegraphics{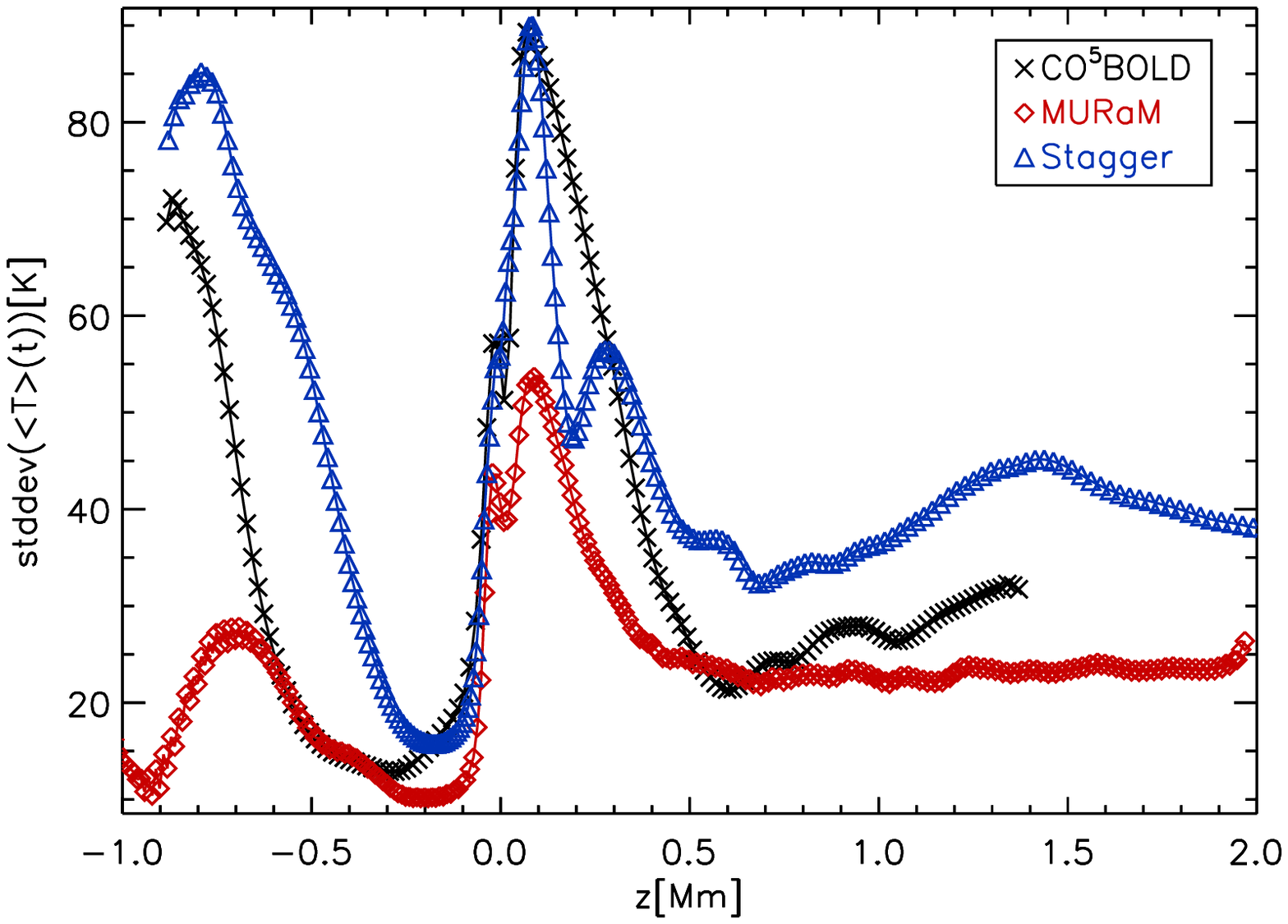}}
  \resizebox{0.8\hsize}{!}{\includegraphics{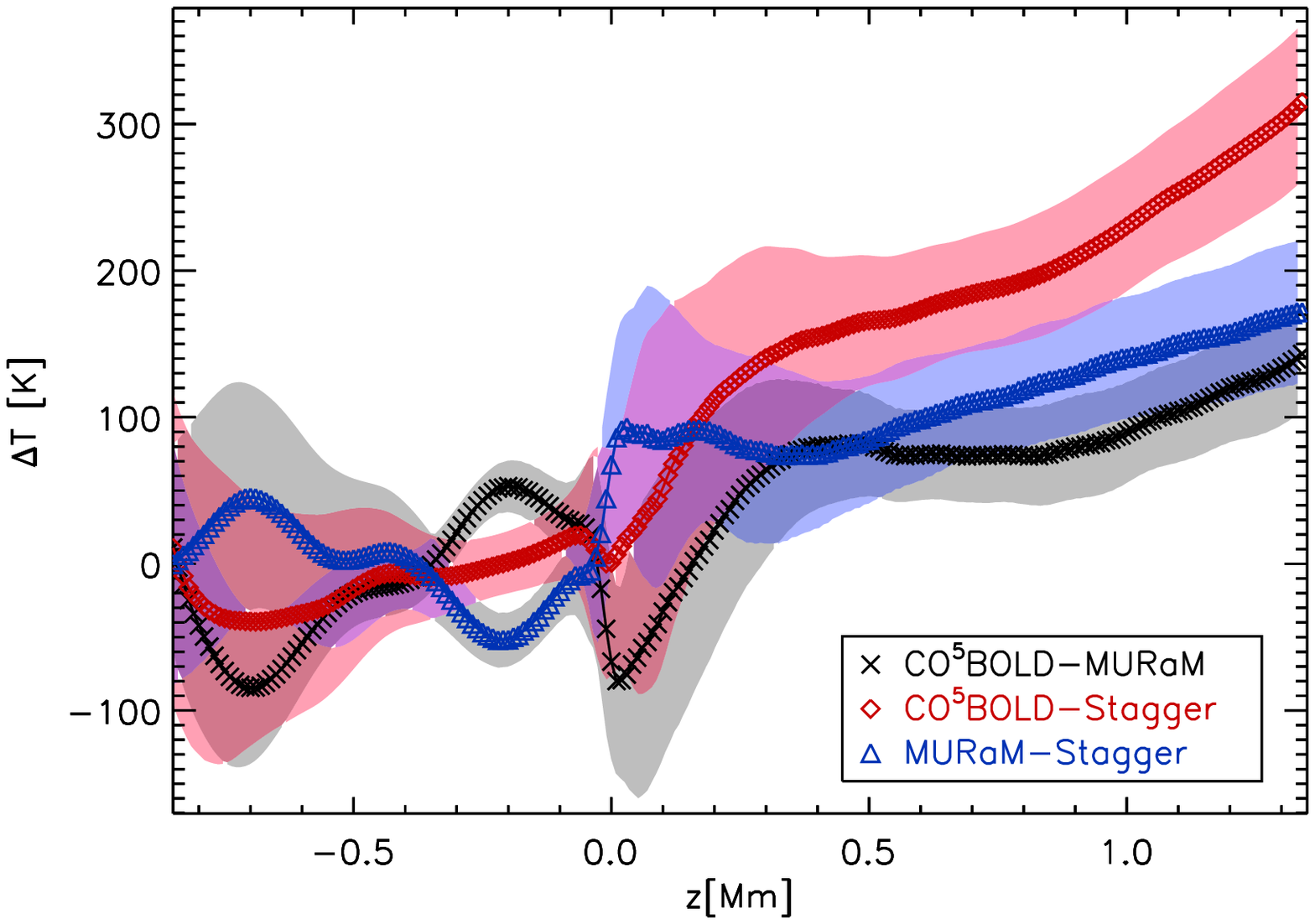}
                          \includegraphics{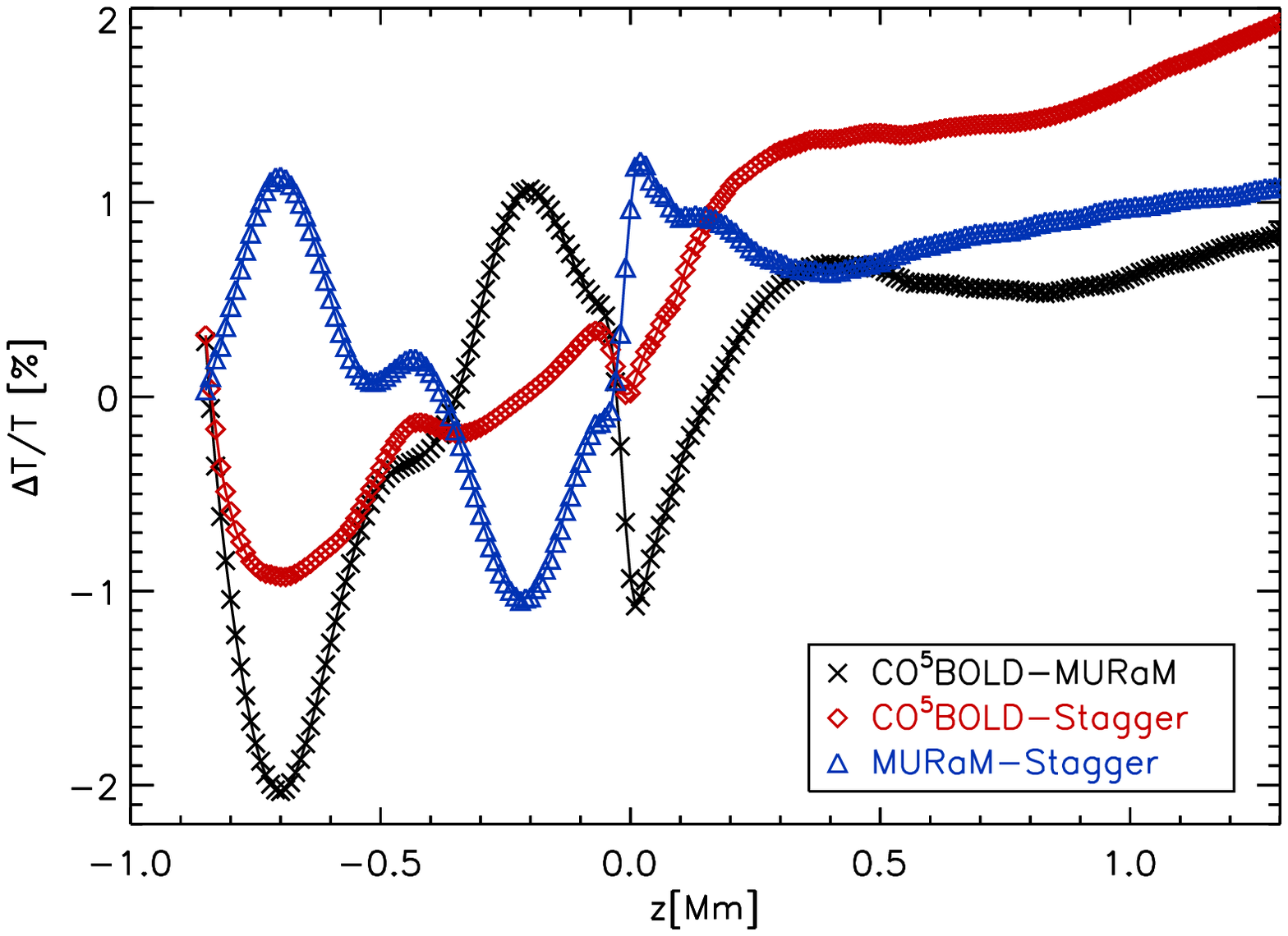}}
\caption{Upper panels: horizontally averaged temperature (left) and
         standard deviation of the mean temperature profiles
         corresponding to the 19 simulation snapshots (right) as
         functions of geometrical depth ($z=0$: average depth of
         $\tau_{500}=1$).  Lower panels: absolute (left) and relative
         (right) mean temperature differences
         between the models.}
\label{fig:t_all}
\end{figure*}
\begin{figure*}[ht!]
\vspace*{0mm}
\centering
  \resizebox{0.8\hsize}{!}{\includegraphics{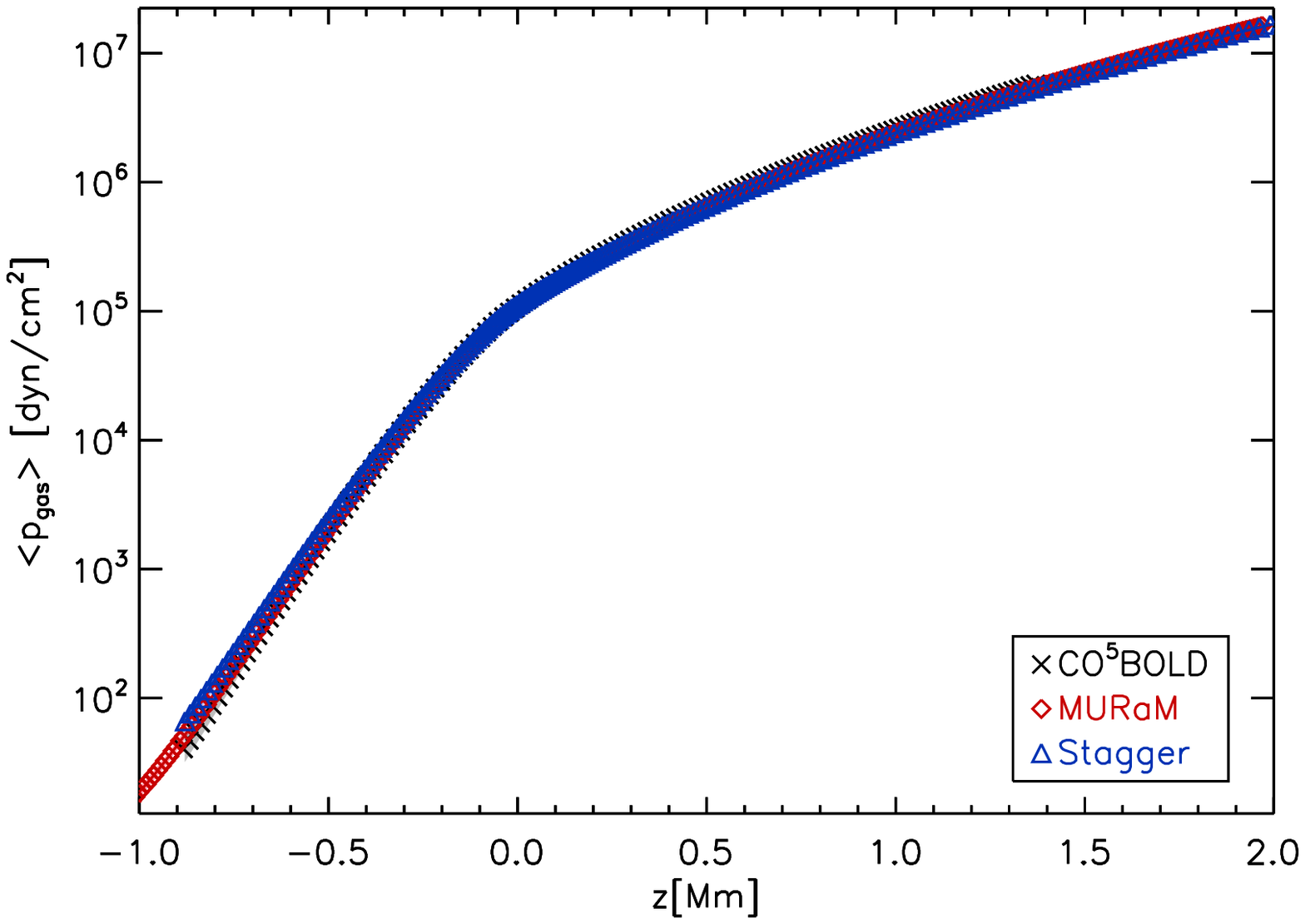}
                          \includegraphics{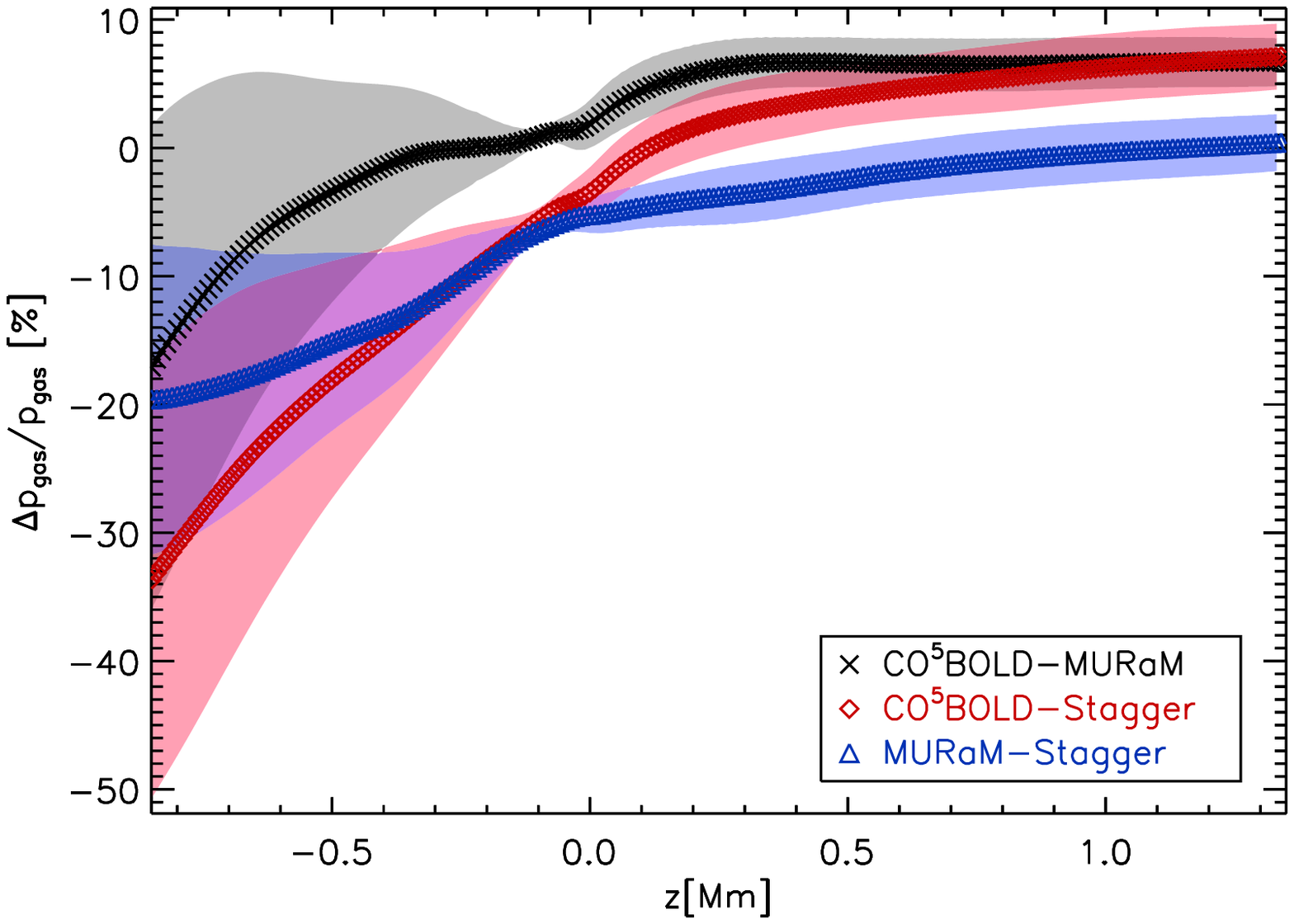}}
  \resizebox{0.8\hsize}{!}{\includegraphics{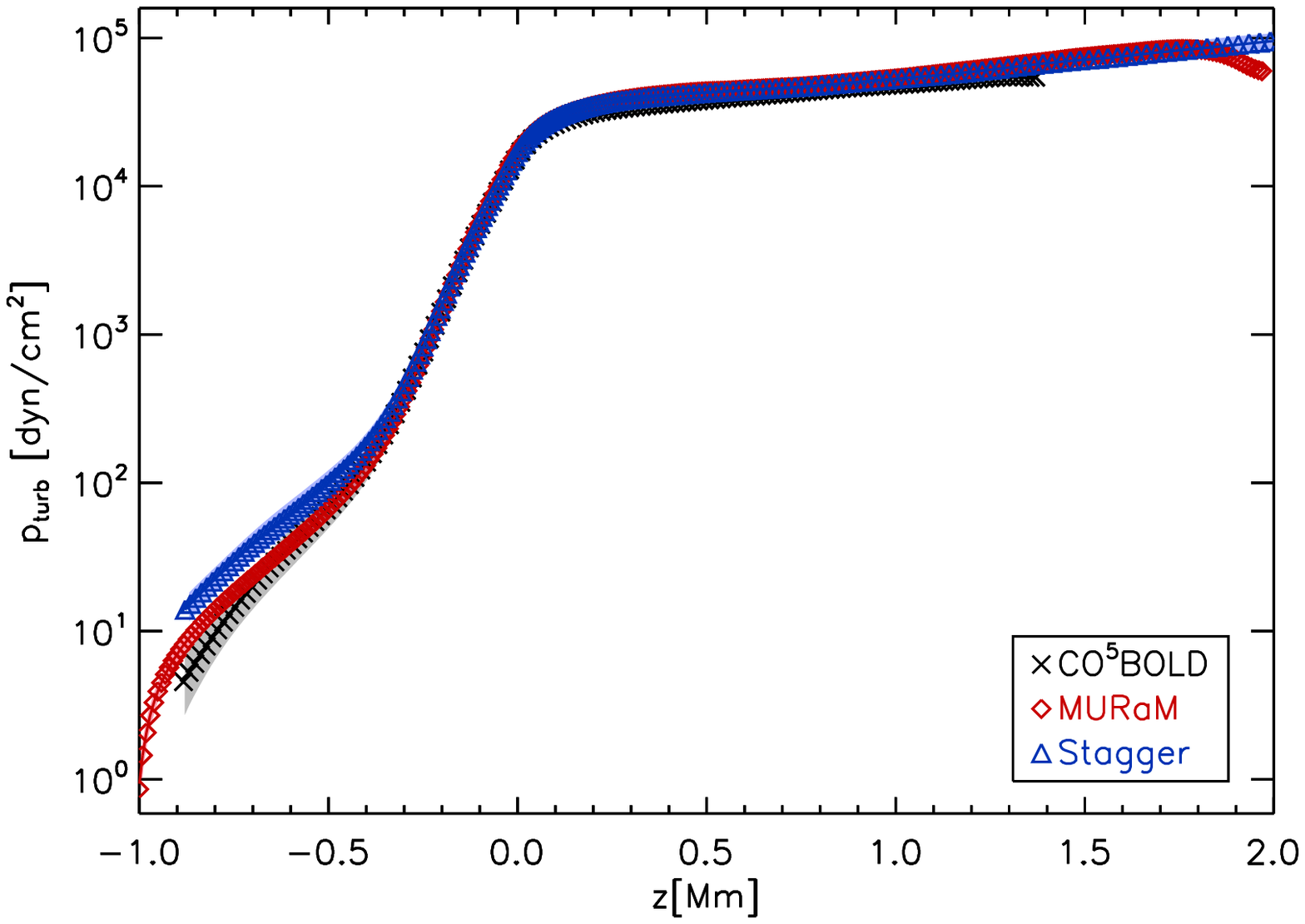}
                          \includegraphics{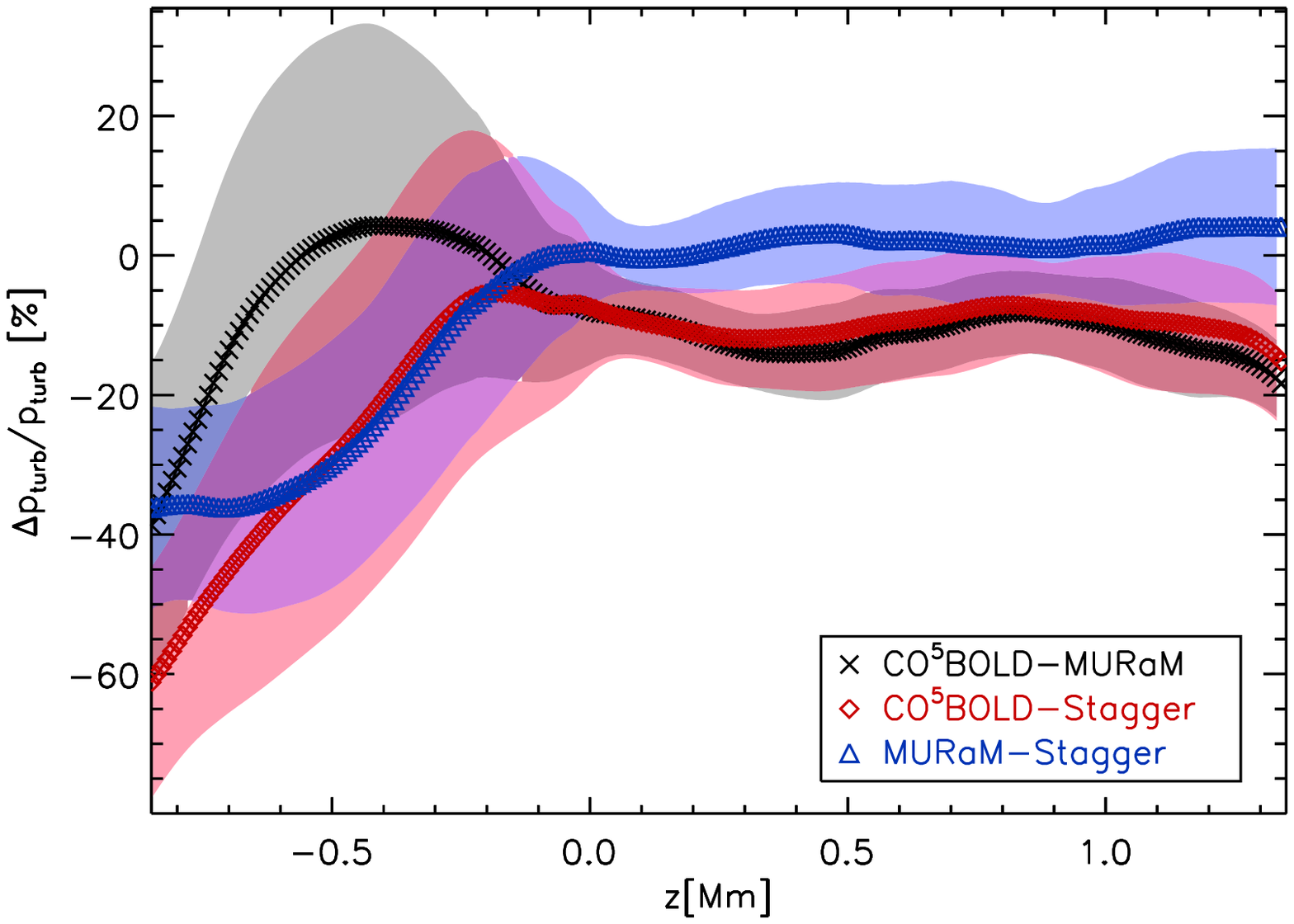}}
\caption{Upper panels: horizontally averaged gas pressure (left) and
         relative differences between the models (right) as functions of
         geometrical depth. Lower panels: same for the horizontally
         averaged turbulent pressure, $p_{\rm turb}= \rho u_z^2$.}
\label{fig:p_all}
\end{figure*}

\subsection{Mean stratification}

The upper panels of Fig.~\ref{fig:t_all} show the geometrical-depth
profiles of the horizontally averaged temperature and their (temporal)
standard deviations (see Eq.~\ref{eq_stddev}), the latter indicating the
level of fluctuations of the mean profiles among the 19 snapshots used
from each of the three simulation runs. The depth scales of the three
models were aligned such that $z=0$ always refers to the average depth
of the surface $\tau_{500}=1$. The weaker fluctuations between the
\MURaM snapshots compared to the other models are probably a result of the
horizontally more extended computational box and the explicit damping of
the fundamental box oscillation mode in this simulation.
\begin{figure}[ht!]
\vspace*{0mm}
\centering
  \resizebox{0.85\hsize}{!}{\includegraphics{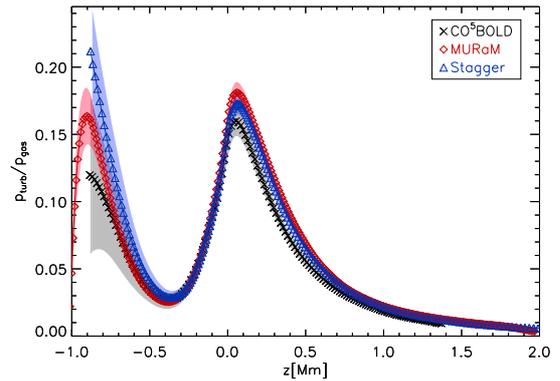}}
\caption{Horizontally averaged ratio of turbulent pressure to gas
         pressure.}
\label{fig:pressure_ratio}
\end{figure}
\begin{figure*}[ht!]
\vspace*{0mm}
\centering
  \resizebox{0.9\hsize}{!}{\includegraphics{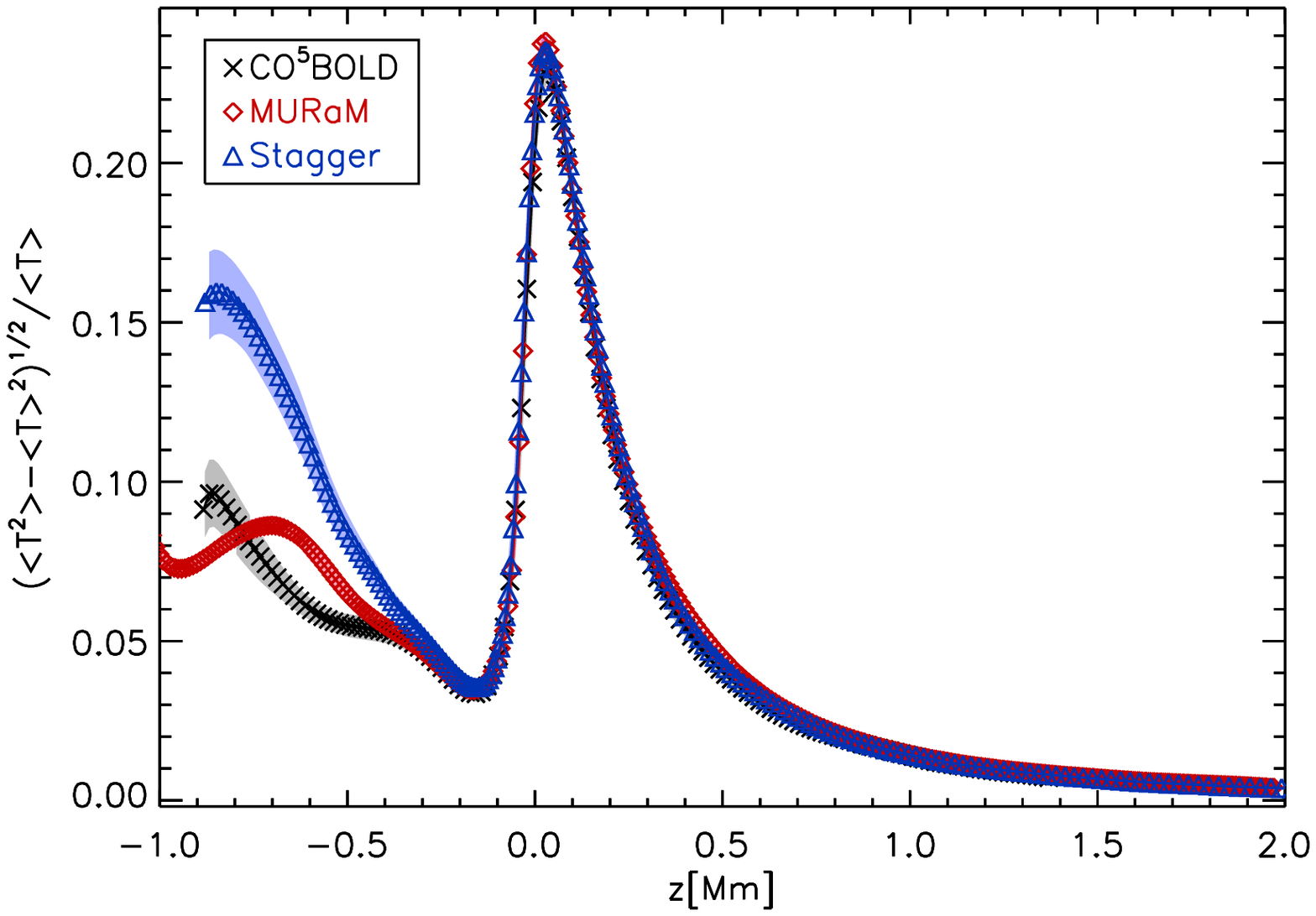}
                          \includegraphics{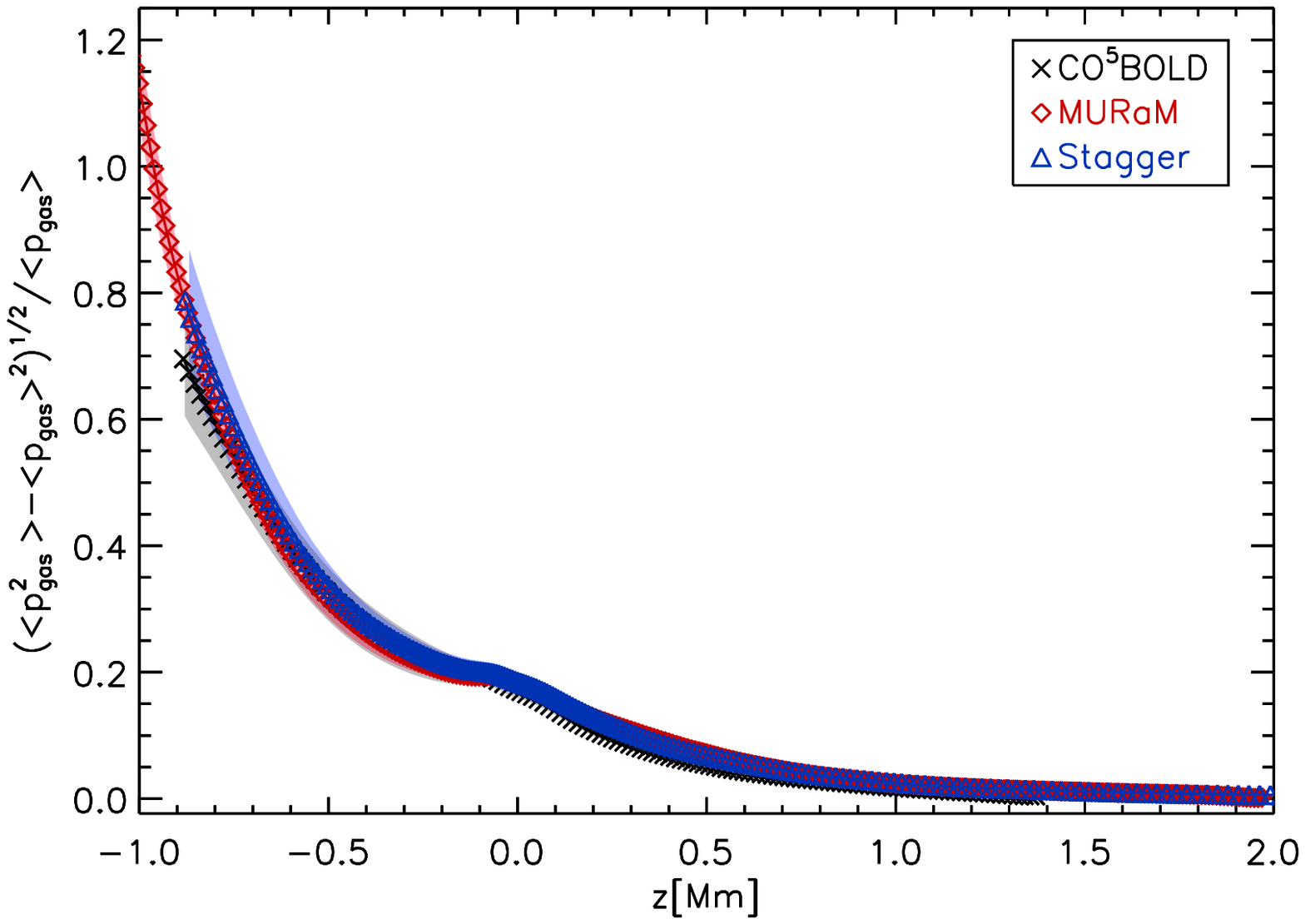}}
\caption{RMS fluctuations of temperature (left) and pressure (right) on
         surfaces of constant geometrical depth. }
\label{fig:rms_T_P_z}
\end{figure*}

\begin{figure*}[ht!]
\vspace*{0mm}
\centering
  \resizebox{0.9\hsize}{!}{\includegraphics{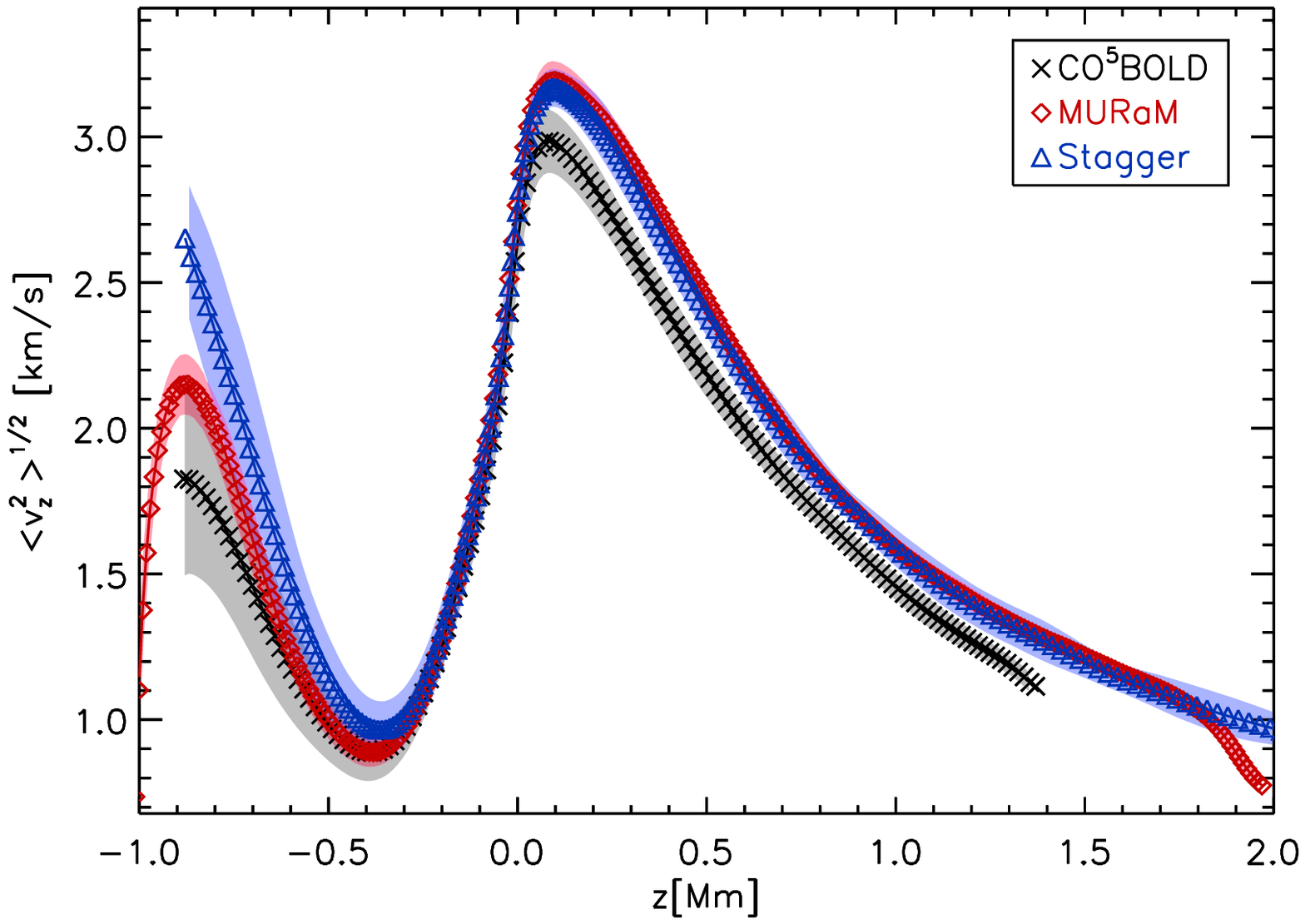}
                          \includegraphics{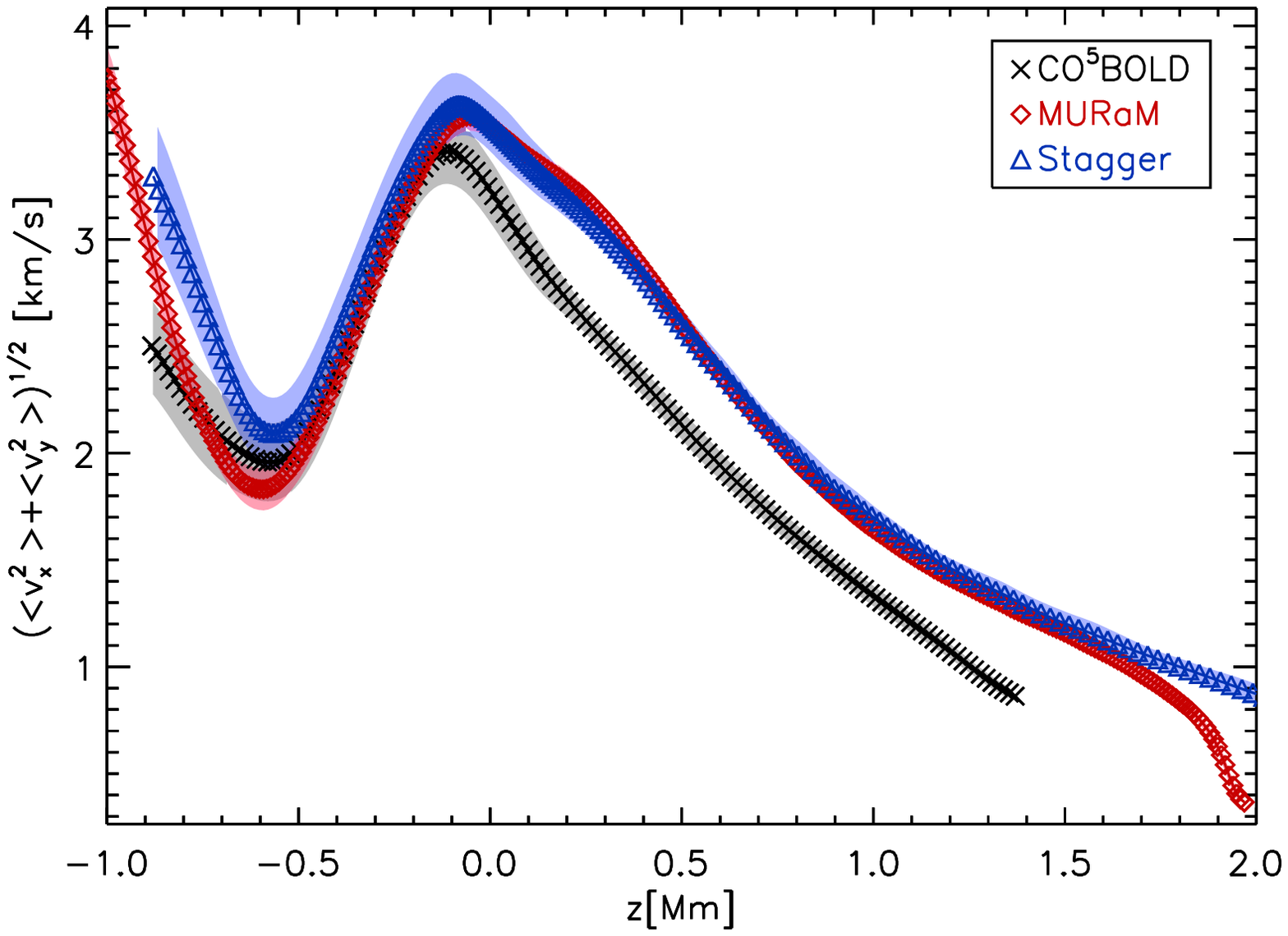}}
\caption{RMS of the vertical (left) and horizontal (right)
         velocity on surfaces of constant geometrical depth.}
\label{fig:rms_vels_z}
\end{figure*}

\begin{figure*}[ht!]
\vspace*{0mm}
\centering
  \resizebox{0.9\hsize}{!}{\includegraphics{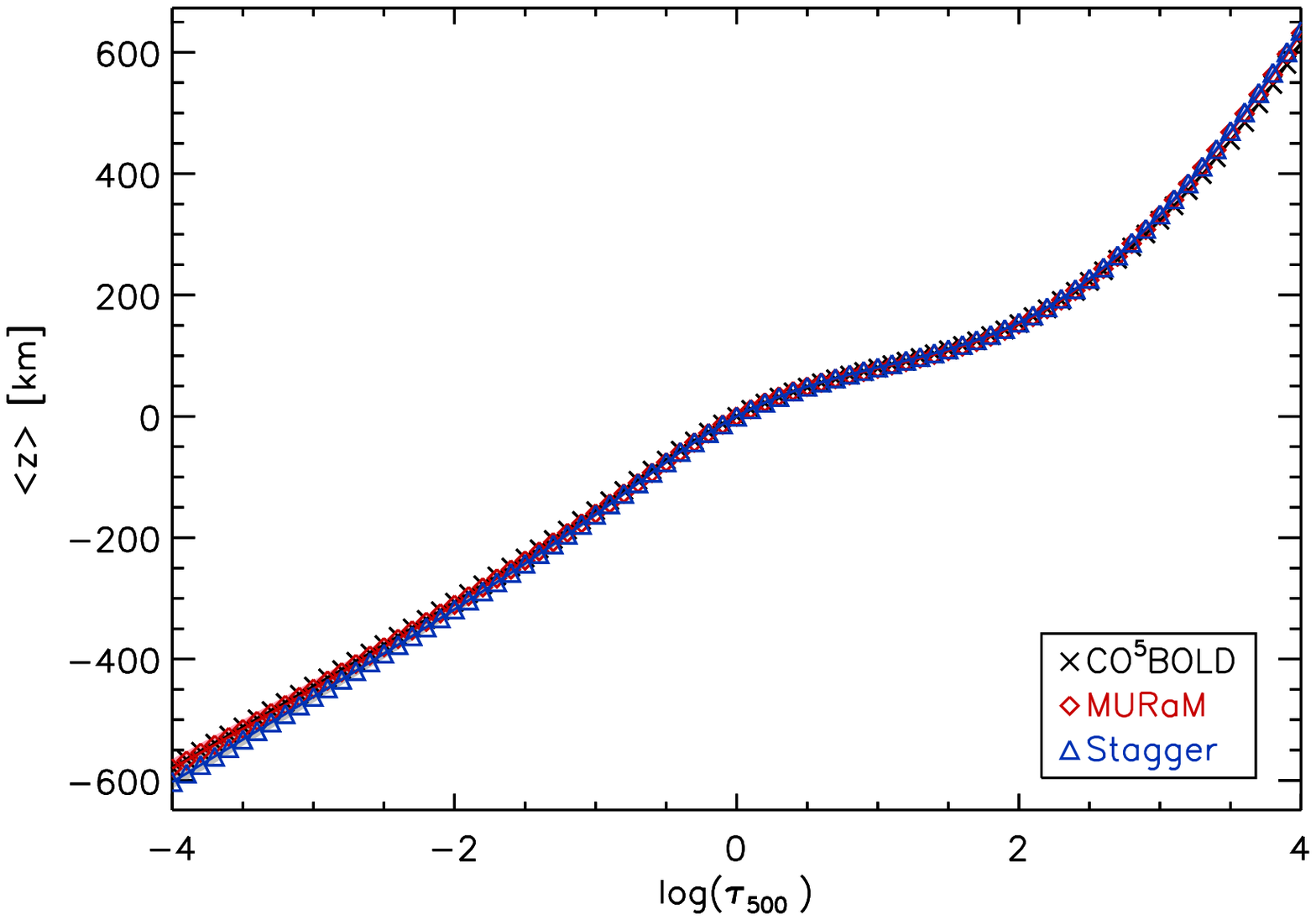}
                          \includegraphics{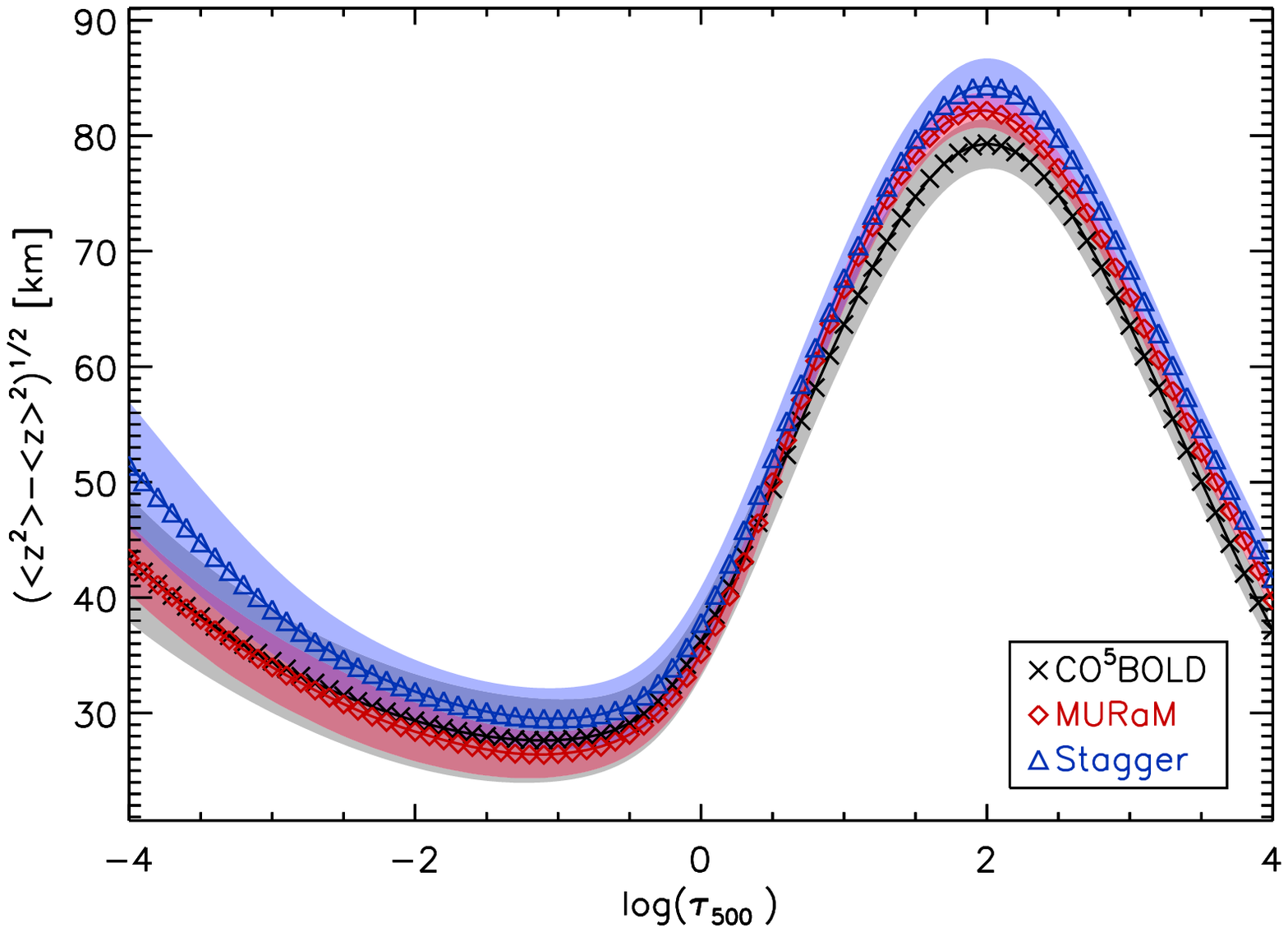}}
\caption{Left: geometrical depth averaged over iso-$\tau$ surfaces
  (left) as functions of continuum optical depth $\tau_{500}$. Right:
  spatial RMS fluctuations of the geometrical depth of the iso-$\tau$
  surfaces.}
\label{fig:z_upper}
\end{figure*}
\begin{figure*}[ht!]
\vspace*{0mm}
\centering
  \resizebox{0.9\hsize}{!}{\includegraphics{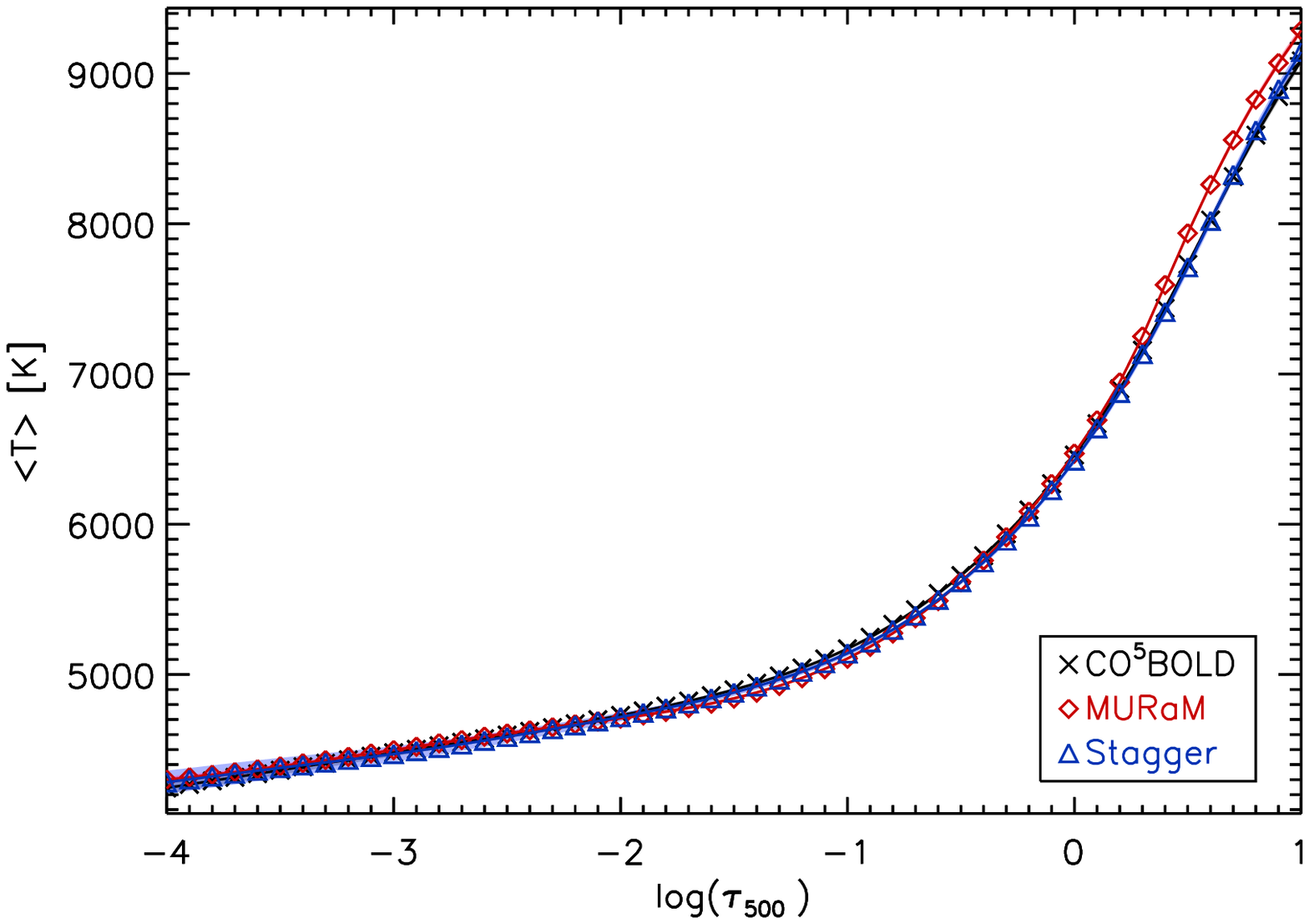}
                          \includegraphics{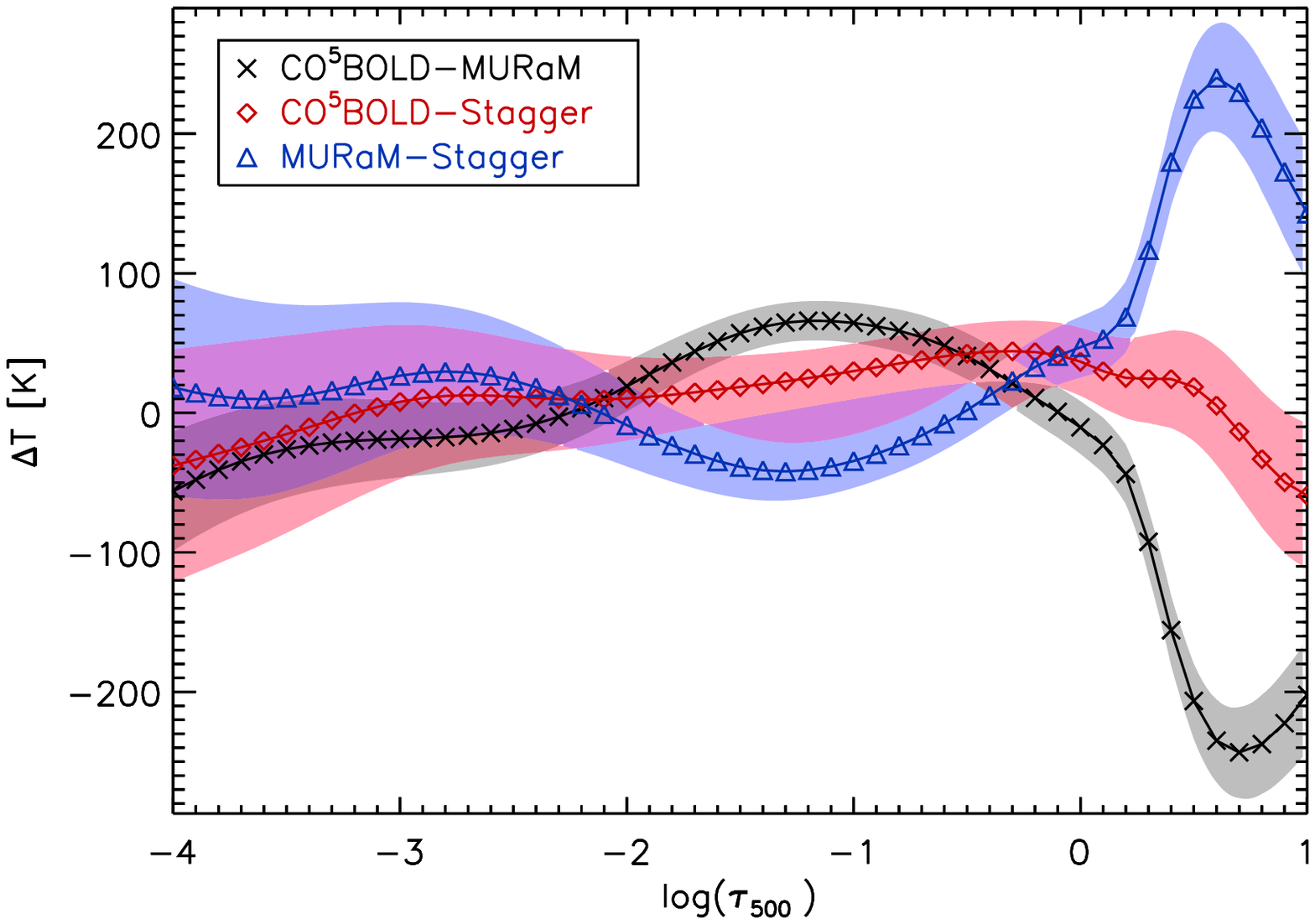}}
\caption{Temperature averaged over iso-$\tau$ surfaces (left) and
         temperature difference between the models (right) as a function
         of optical depth $\tau_{500}$ in the photospheric layers.}
\label{fig:t_upper}
\end{figure*}

The absolute and relative differences between the mean temperature
profiles from the three simulations are given in the lower panels of
Fig.~\ref{fig:t_all}.  The colored bands in this plot (and in similar
subsequent figures) represent the sample standard deviations%
\footnote{The standard deviation of the differences is equal to the
square root of the sum of the variances according to the 19 snapshots of
the individual models.}
of the differences between pairs of models on the basis of the 19
snapshots from each simulation, indicating the range of
scatter caused by the temporal variability of the averages in the
relatively small simulation boxes.  The temperature profiles agree
fairly well, with relative differences below 2\% and
overlapping bands of the standard deviation over most of the height
range in the lower left panel of Fig.~\ref{fig:t_all}. The positive and
negative excursions near $z=0$ result from slight differences in the
depth location and slope of the steep temperature gradient near optical
depth unity. These differences can easily arise given the strong
temperature dependence of the continuum opacity around
$T(\tau_{500}=1)\simeq 6400$~K and the differing vertical resolution of
the simulations. In the photosphere ($-500\,$km$\le z \le 0\,$km), the
\COBOLD and \STAGGER profiles deviate by less than 20~K from each other
while the \MURaM temperatures differ somewhat more, up to 60~K (i.e., at
a level of about 1\%).

Figure~\ref{fig:p_all} shows the depth profiles of gas pressure, $p_{\rm
gas}$, and of turbulent pressure, $\langle\rho v_z^2\rangle$, together
with the respective relative differences between the simulations.  The
ratio of turbulent to gas pressure is given in
Fig.~\ref{fig:pressure_ratio}. The turbulent pressure reaches 
nearly 20\% of the gas pressure slightly below the optical surface
(where the vertical velocity fluctuations peak) and again about this
value in the top layers.

\begin{figure*}[ht!]
\vspace*{0mm}
\centering
  \resizebox{0.9\hsize}{!}{\includegraphics{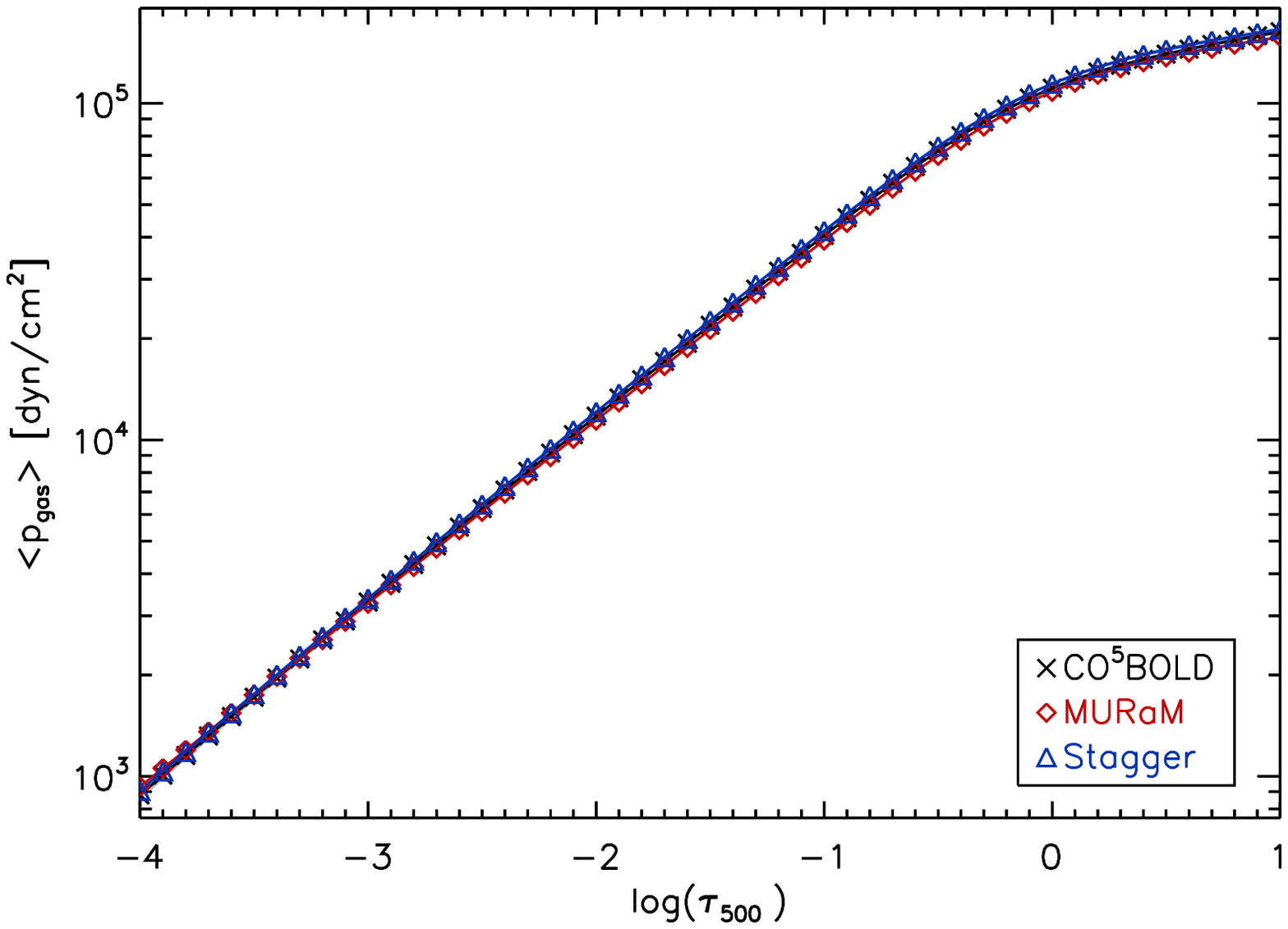}
                          \includegraphics{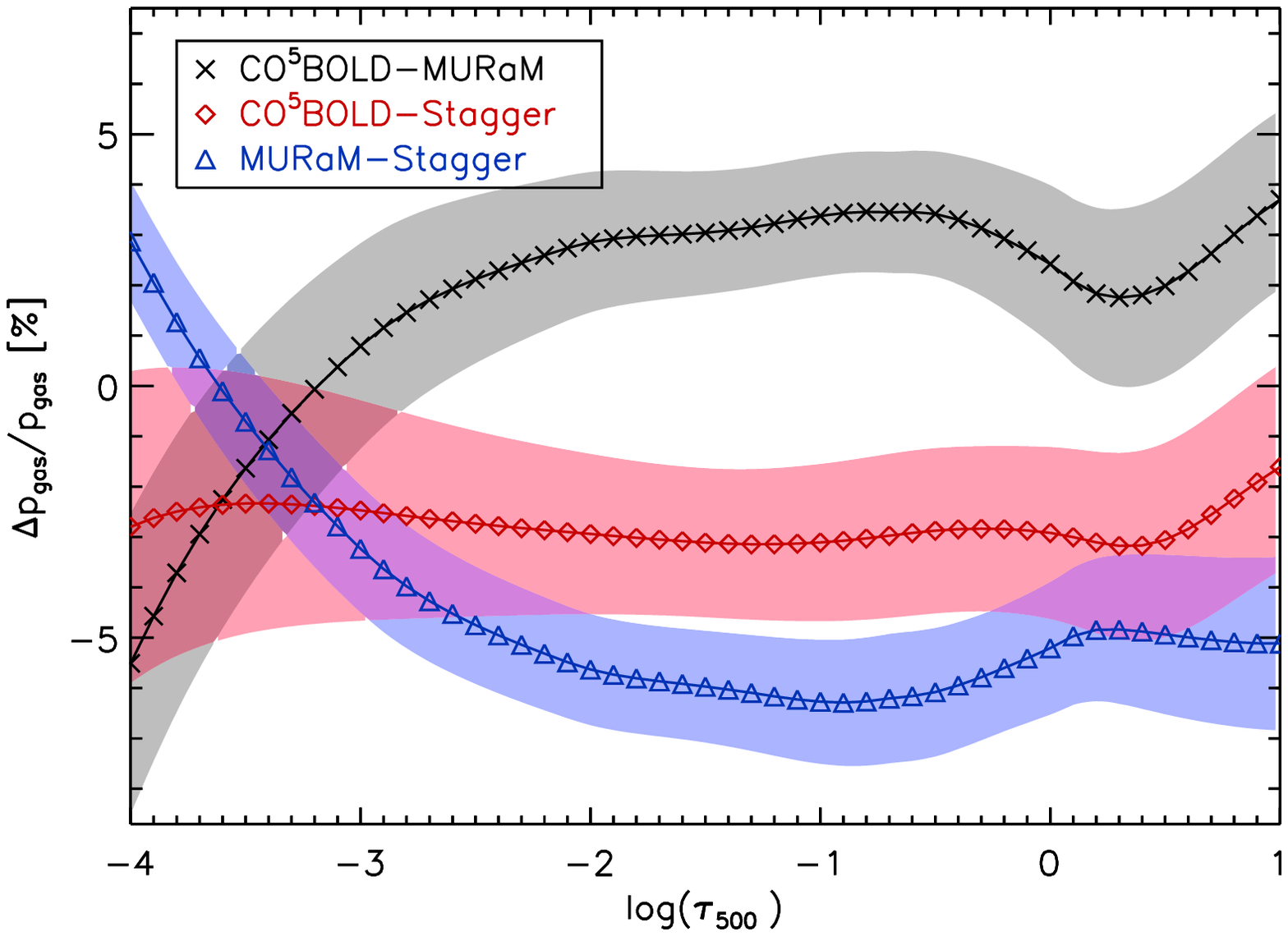}}
\caption{Gas pressure profiles (left) and relative differences
         (right) averaged over iso-$\tau$ surfaces as functions of
         optical depth $\tau_{500}$. }
\label{fig:p_upper}
\end{figure*}

The relative differences of the gas pressure between the models are
somewhat larger than those of the temperature, especially in the
uppermost layers. While the roughly depth-independent deviations in the
layers below $z=0$ can be simply explained by a constant relative shift
of the respective geometrical height scales, the bigger deviations in
the photosphere are caused by the significant differences of the
turbulent pressure in these layers between the simulations (cf. the
lower right panel of Fig.~\ref{fig:p_all}). To illustrate the effect,
consider the simple case of an isothermal atmosphere and constant
turbulent speed, $v_z$. The scale height of the gas pressure is then
given by $H_p=(c_{\rm s}^2+v_z^2)/g$, where $c_{\rm s}$ is the sound
speed and $g$ is the (constant) gravitational acceleration.  If the
turbulent pressure differs between two stratifications, the effect is
cumulative: for instance, a difference of 5\% over 6 scale heights adds
up to 30\% of a scale height, leading to a significant pressure
deviation in the upper layers. The higher turbulent speeds of the
\STAGGER model imply a larger scale height and, therefore, higher
pressure and density in the upper layers. In terms of optical depth
(relevant for observations), the deviations of the pressure
stratifications are significantly smaller (cf. Fig.~\ref{fig:p_upper})
since higher values of pressure and density lead to an upward shift of
the iso-$\tau$ surfaces (and vice versa).

Figure~\ref{fig:rms_T_P_z} shows the relative RMS fluctuations of
temperature and pressure, respectively, on surfaces of constant
geometrical depth (see Eq.~\ref{eq_rms}). The temperature fluctuations
(left panel) show a sharp maximum near the optical surface, where
radiative cooling leads to strong temperature differences between
granular upflows and intergranular downflow regions. The development of
shocks in the uppermost layers of the simulation boxes results in a
second peak of the temperature fluctuations. In contrast, the pressure
fluctuations grow monotonically outward and reach their maximum at
the top of the simulated regions.

The RMS values of the velocity components are shown in
Fig.~\ref{fig:rms_vels_z}. The RMS of the vertical velocity (left panel)
peaks near the optical surface, owing to the braking of the upflows and
the acceleration of the cool downflows as a result of radiative
cooling. Because the scale height decreases rapidly, most of the rising
fluid that reaches the surface has to overturn very near to optical
depth unity, so that the RMS value of the horizontal velocity (right
panel of Fig.~\ref{fig:rms_vels_z}) peaks only slightly higher than
those of the vertical velocity. The RMS of both velocity components grow
again in the upper, shock-dominated top layers of the simulation
boxes. The somewhat lower RMS values of the \COBOLD model are probably
related to the shallower computational box in comparison to the other
simulations.  The drop near the bottom boundary in the case of the
\MURaM simulation is caused by a narrow layer of enhanced viscosity,
which was introduced on grounds of numerical stability. In spite of
these differences, all three codes show excellent agreement in the
observable photospheric layers ($-0.5\,$Mm$\,\, < z < 0\,$Mm).

\begin{figure*}[ht!]
\vspace*{0mm}
\centering
  \resizebox{0.9\hsize}{!}{\includegraphics{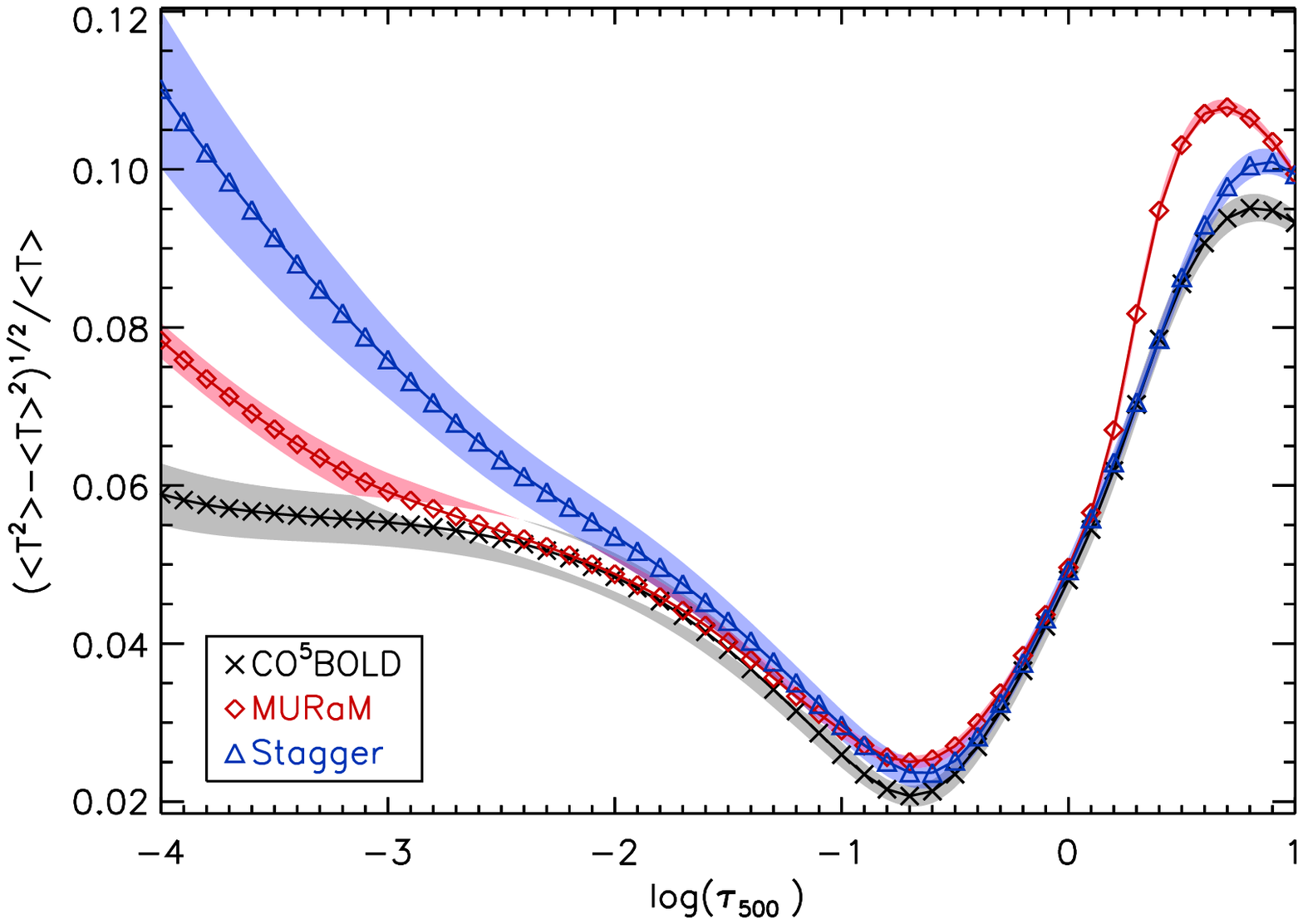}
                          \includegraphics{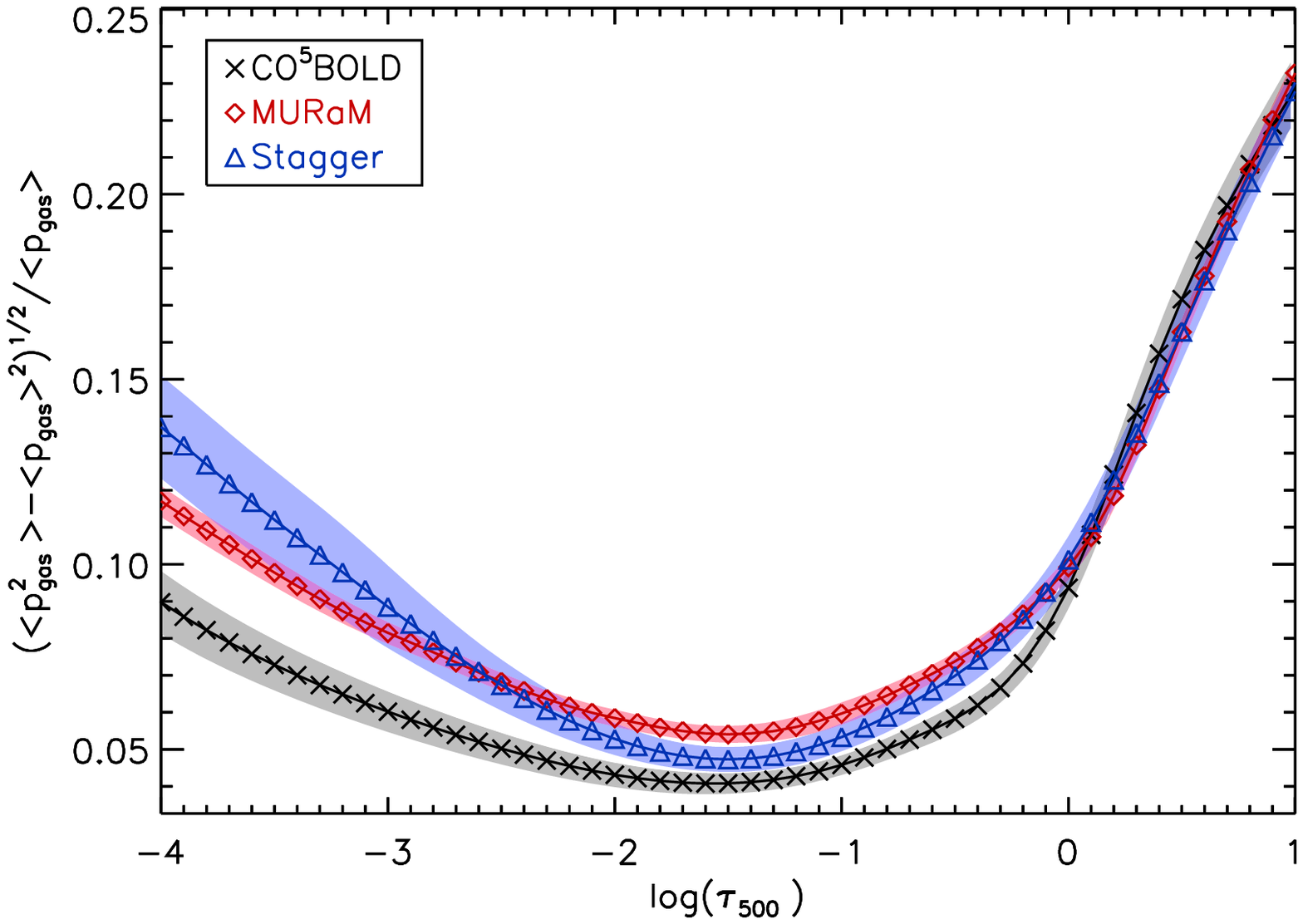}}
\caption{Relative RMS fluctuations of temperature (left) and pressure
         (right) on surfaces of constant optical depth $\tau_{500}$. }
\label{fig:rms_T_P_tau}
\end{figure*}

\subsection{Photospheric structure}

For all observables originating in the solar photosphere, the profiles
of the physical quantities as functions of optical depth are relevant.
Since the surfaces of constant optical depth are not flat but strongly
corrugated in the photosphere owing to granulation, the profiles of
quantities averaged over surfaces of constant optical depth generally do
not simply correspond to stretched or shifted profiles of horizontally
averaged quantities.

Figure~\ref{fig:z_upper} illustrates the relation between the
geometrical and the optical depth scales. The left panel shows the mean
geometrical depths of the iso-$\tau$ surfaces as a function of continuum
optical depth at 500~nm. The profiles are very similar: the maximum
differences between the models are on the order of the vertical
distances of the grid cells. The (spatial) RMS fluctuations of the depth
of the iso-$\tau$ surfaces are shown in the left panel of
Fig.~\ref{fig:z_upper}. They quantify the `corrugation' of the
iso-$\tau$ surfaces, which reaches a maximum of $\sim$$80\,$km somewhat
below the optical surface, presumably because of the dominant effect of the
cool downflow regions and the strong (positive) temperature dependence
of the H$^-$ continuum opacity.

The profiles of temperature averaged over iso-$\tau$ surfaces and the
differences between the models are shown in Fig.~\ref{fig:t_upper}. In
the photosphere, the biggest difference between the models appears in
the range $-1.5 < \log\tau_{500} < -0.5$, where the \MURaM model is up
to 40~K cooler than the \STAGGER model and 65~K cooler ($\sim 1$\%) than
the \COBOLD model.  This probably is a result of the less detailed
opacity binning procedure used in the \MURaM simulation (only 4 opacity
bins compared to 12 in the other simulations).  The deviation leads to a
somewhat lower temperature gradient in the \MURaM case between $-2.5 <
\log\tau_{500} < -1.5$. In a narrow layer 
around $\log\tau_{500} \simeq 0.7$ below the optical surface, the \MURaM
model is up to 200~K hotter than the other two models. This could be
related to the higher spatial resolution of the \MURaM simulation, which
leads to less artificial diffusion of heat between the hot upflows and
cool downflows, thus maintaining bigger temperature fluctuations 
(cf. Fig.~\ref{fig:rms_T_P_tau}).

The profiles of gas pressure averaged over iso-$\tau$ surfaces are shown
in Fig.~\ref{fig:p_upper}. The relative differences
between the models in the photosphere generally amount to a few percent.
The biggest differences of up to 6.5\% arise between the \STAGGER and
\MURaM models near $\log\tau_{500}=-1$. As explained in the preceding
subsection, these differences are related to the effect of the turbulent
pressure on the pressure scale height.

Figure~\ref{fig:rms_T_P_tau} shows the spatial RMS fluctuations of
temperature and pressure on iso-$\tau$ surfaces. In the range $-2 \leq
\log\tau_{500} \leq 0$, the horizontally averaged temperature
fluctuations are similar for all models.  In the higher layers,
the models deviate more strongly from each other. This is a result of
the differences in the velocity fluctuations and turbulent pressure
discussed above.

\begin{figure*}[ht!]
\vspace*{0mm}
\centering
  \resizebox{1.0\hsize}{!}{\includegraphics{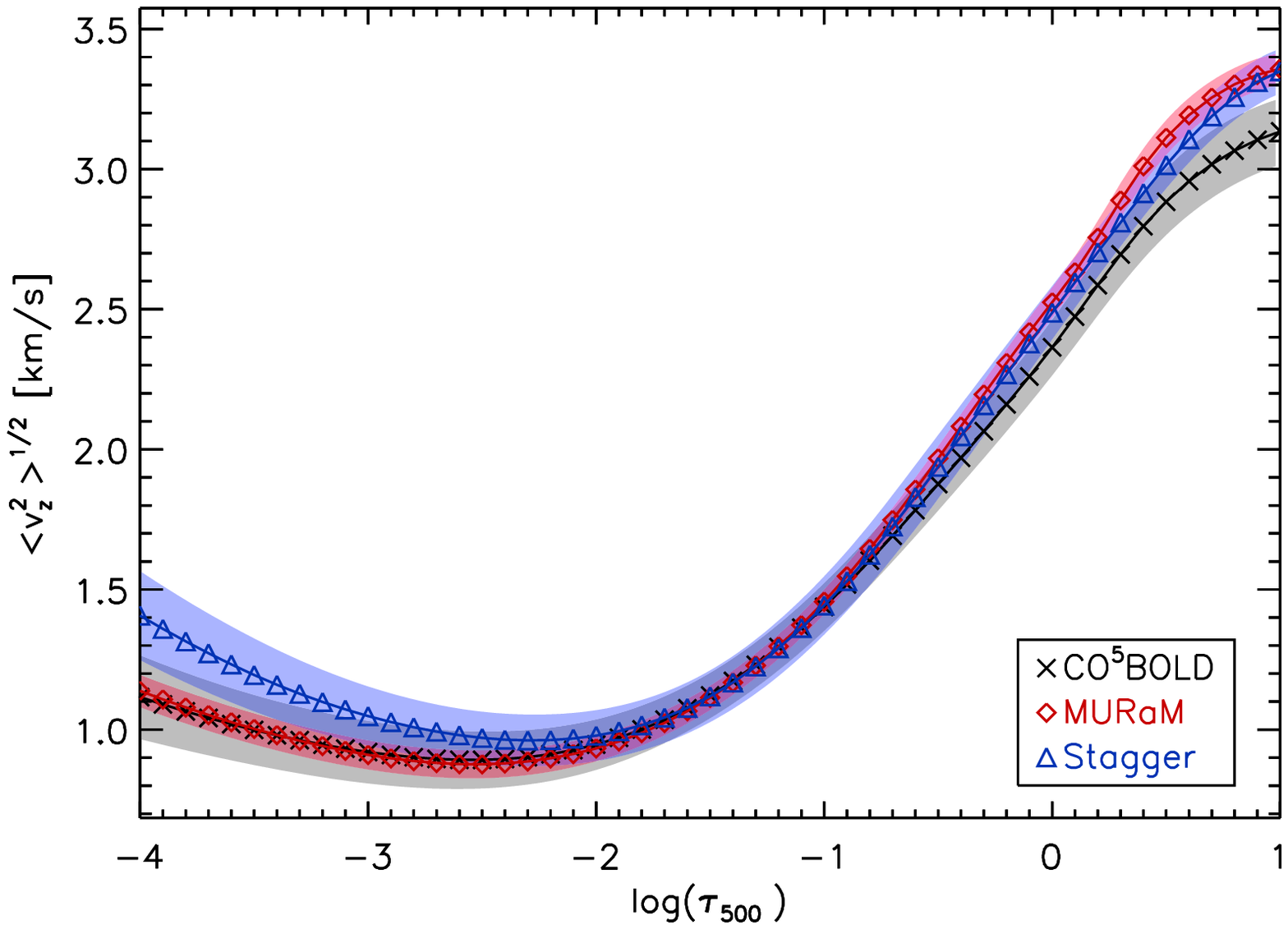}
                          \includegraphics{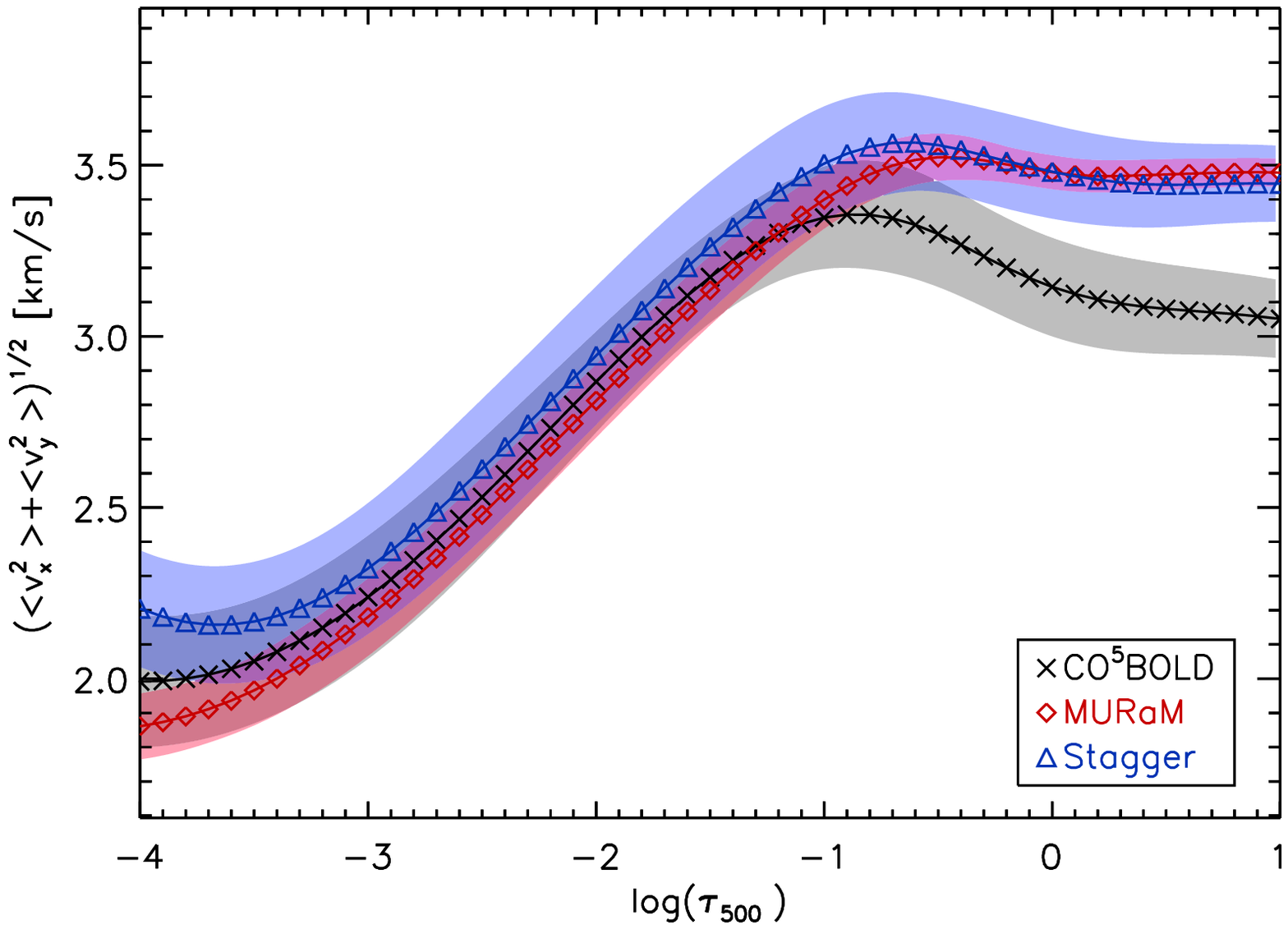}}
\caption{RMS of the vertical (left) and horizontal (right)
         velocity on surfaces of constant optical depth $\tau_{500}$.}
\label{fig:rms_vels_tau}
\end{figure*}

The RMS values of the vertical and horizontal components of the fluid
velocity are shown in Fig.~\ref{fig:rms_vels_tau}.  The RMS values of
both components in the photosphere are very similar for all three
simulations, while the deviations become somewhat larger in the layers
above.

\begin{figure*}[ht!]
\vspace*{0mm}
\centering
  \resizebox{1.0\hsize}{!}{\includegraphics{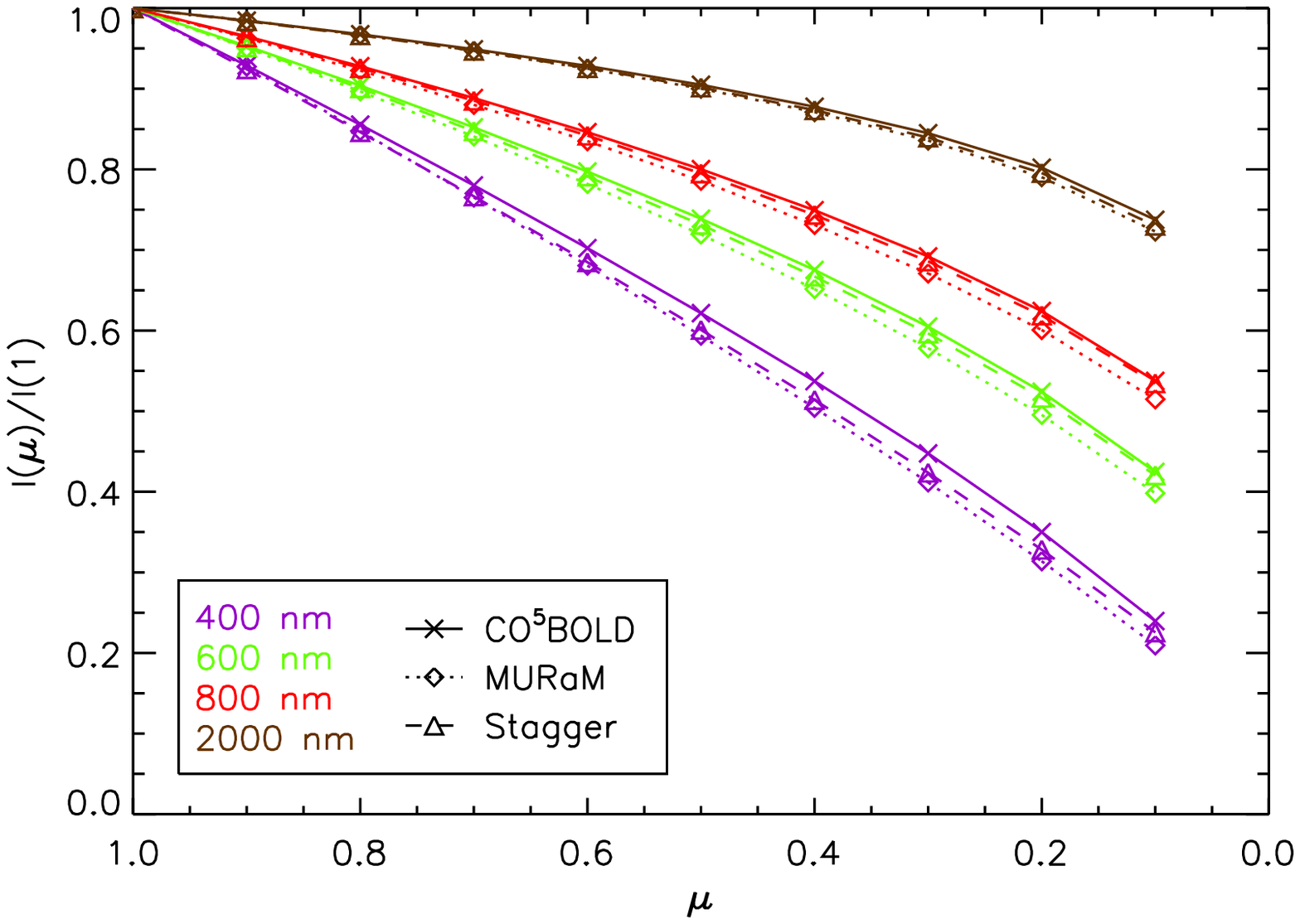}
                          \includegraphics{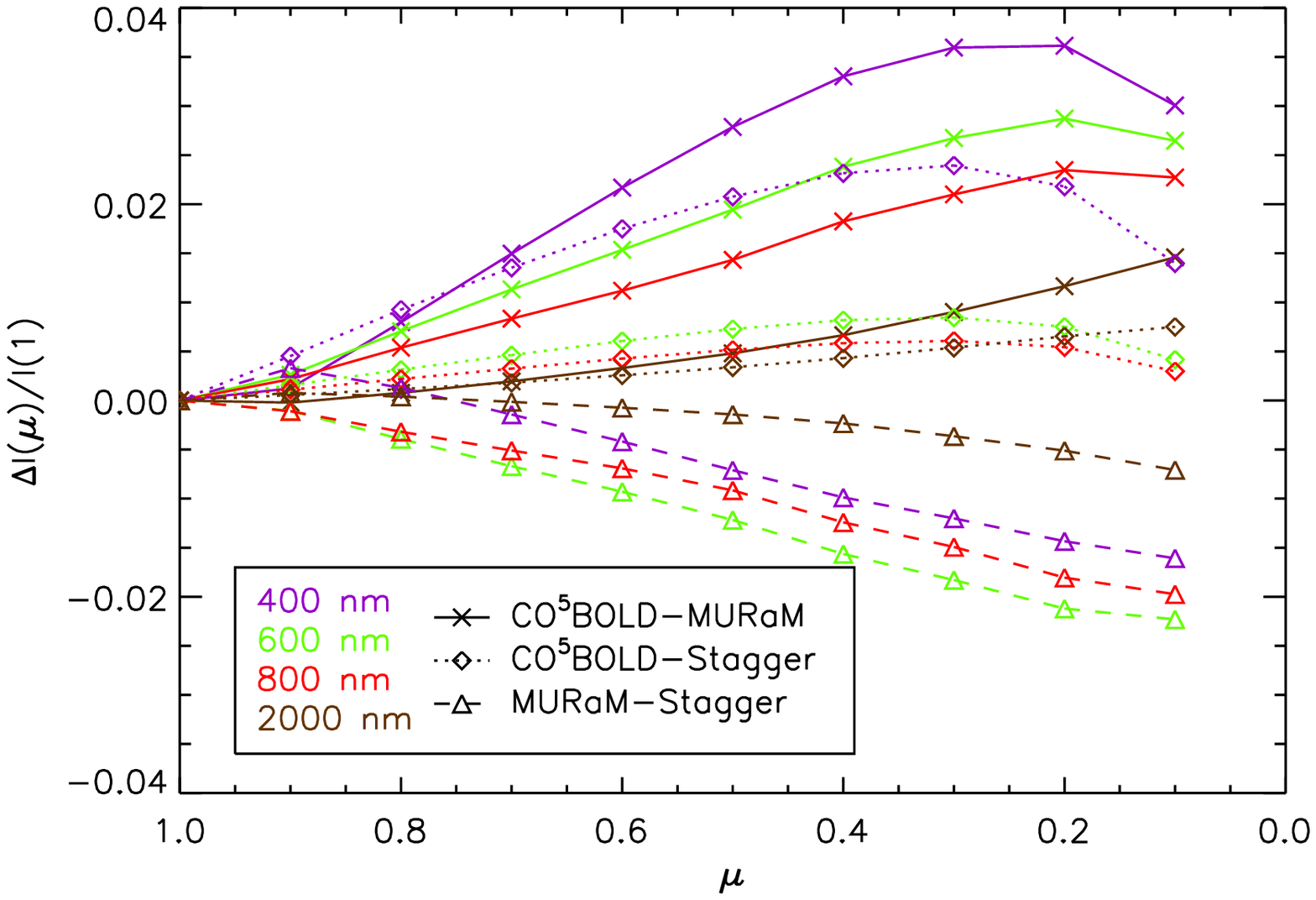}}
\caption{Center-to-limb variation of the continuum intensity (left) and
  differences between the models (right) as a function of
  $\mu=\cos\theta$ at four wavelengths in the visible and infrared
  spectral ranges.}
\label{fig:clv}
\end{figure*}

\subsection{Center-to-limb variation of continuum intensity}
In order to estimate how much the differences between the models affect
observable quantities, we considered the center-to-limb variation (CLV)
of the monochromatic continuum intensity. The intensities were
calculated at four wavelengths ($400\,$nm, $600\,$nm, $800\,$nm, and
$2000\,$nm) along rays with 10 inclination angles, with $\mu=\cos\theta$
ranging from $\mu=1$ (disk center) to $\mu=0.1$ (limb). For each
inclination and wavelength, the intensities were spatially averaged over
the simulated surface areas, temporarily averaged over simulation
snapshots, and azimuthally averaged over four (\COBOLD, \MURaM) or 12
(\STAGGER) equidistant azimuthal directions.

The CLVs for the \COBOLD simulation were computed with the spectrum
synthesis code \texttt{Linfor3D}%
\footnote{http://www.aip.de/$\sim$mst/Linfor3D/linfor\_3D\_manual.pdf}.
The NLTE EOS used in Linfor3D is more detailed than that employed in
\COBOLD, because it has to provide the electron pressure and the number
density for all individual atoms and ions. Likewise, the continuum
opacities used in \texttt{Linfor3D} are not fully consistent with the
raw opacities from which the binned opacities in the \COBOLD simulations
are constructed. However, both opacities are based on the same chemical
abundance mix \citep[][with the exception of CNO, for which the values
A(C) = 8.41, A(N) = 7.80, A(O) = 8.67 are adopted]
{Grevesse+Sauval:1998}.  The emergent continuum intensity is obtained by
integrating the transfer equation on a grid that is refined with respect
to the original hydrodynamics grid, ensuring that the resolution in
vertical optical depth is about $\Delta \tau_{\rm Ross} \approx 0.1$.

The CLVs for the snapshots of the \STAGGER simulation were computed with
the line formation code \texttt{SCATE} \citep{Hayek:2011}, using the
same Feautrier-like long-characteristics solver as in the original
simulation. \texttt{SCATE} employs the same EOS as the
\STAGGER code simulation and the same continuous opacities used for the
opacity binning.

For the \MURaM simulation, the CLVs were calculated using a
long-characteristics scheme with automatic grid refinement in case of
steep gradients in optical depth along the ray. It uses the same EOS as
the \MURaM simulation and the same continuum opacities as employed for
the opacity binning \citep[based on the opacity distribution functions
from the ATLAS9 package, see][]{Kurucz:1993}. The continuum opacities
are averages over $20\,$nm (for the CLVs at $400\,$nm, $600\,$nm, and
$800\,$nm) or over $90\,$nm (for $2000\,$nm).

Fig.~\ref{fig:clv} shows the resulting profiles for the three
simulations and the differences between them. The profiles are very
similar, the agreement between \COBOLD and \STAGGER simulations being
somewhat better (except at 400~nm) than that between \MURaM and the
other simulations. The somewhat stronger limb darkening mainly results
from the slightly steeper temperature gradient in the lower
photosphere of the \MURaM model (cf. Fig.~\ref{fig:t_upper}).

\section{Conclusions}
Although the three numerical solar models considered here (\COBOLD,
\MURaM, and \STAGGER) result from codes with different numerical
methods and from simulation boxes of different size and spatial grid
resolution, their overall agreement is very good and encouraging. This
does not only concern the mean (optical and geometrical) depth profiles
of the basic quantities, but also the spatial fluctuations of these
quantities as well as histograms of velocity and intensity, i.e., the
dynamics and spatial structure of the simulated atmospheres. Slight
deviations between the models for some quantities are probably caused by
differences in spatial resolution and in the opacity binning
procedures. These results give confidence in the reliability of
comprehensive simulations as a tool for studying stellar atmospheres and
surface convection.

\begin{acknowledgements}
B.F. acknowledges financial support from the Agence Nationale de
la Recherche (ANR) and the Programme National de Physique
Stellaire (PNPS) of CNRS/INSU, France. W.H. acknowledges support by
the European Research Council under the European Community's 7th
Framework Programme (FP7/2007--2013 Grant Agreement no. 247060).
\end{acknowledgements}

\bibliography{paper_final.bbl}

\end{document}